

\def\ko{K_0}
\def\kob{{\bar K}_0}
\def\kkbar{$\ko$--$\kob$}
\def\cp{{\it CP\/}}
\def\cpt{{\it CPT\/}}
\overfullrule=0pt
\pubnum={6454}
\date={March, 1994; rev. August, 1994}
\pubtype={T/E}
\titlepage
\title{Violation of CPT and Quantum Mechanics in the \kkbar\ System}
\author{Patrick Huet and Michael E. Peskin\doeack}
\SLAC
\abstract{We reconsider the model of quantum mechanics violation in
the $\ko$--$\kob$ system, due to Ellis, Hagelin, Nanopoulos, and
Srednicki, in which \cp- and \cpt-violating signatures arise from the
evolution of pure states into mixed states.  We present a formalism
for computing time-dependent asymmetries in this model and show that
present data constrains its parameters significantly.  In the future,
this model will be put to very stringent tests at a $\phi$ factory.
We present the theory of these tests and show the relation between
particular $\phi$ decay correlations and the parameters of quantum
mechanics violation.}
 \submit{Nuclear Physics B}
\endpage
\def\half{{1\over 2}}
\def\ml{m_L}
\def\ms{m_S}
\def\gl{\Gamma_L}
\def\gs{\Gamma_S}

\def\ek{\epsilon}
\def\dk{\Delta}

\def\gam{\Gamma}
\def\dm{\Delta m}
\def\dg{\Delta \gam}
\def\kl{K_L}
\def\ks{K_S}
\def\rk{\rho_K}

\def\rl{\rho_L}
\def\rs{\rho_S}
\def\ri{\rho_I}
\def\rib{\rho_{\bar I}}
\def\gl{\gam_L}
\def\gs{\gam_S}
\def\\ml{m_L}
\def\ms{m_S}
\def\gb{{\bar \gam}}
\def\mb{{\bar m}}
\def\egl{e^{-\gl\tau}}
\def\egs{e^{-\gs\tau}}
\def\egi{e^{-\gb\tau}}
\def\egic{e^{-(\gb+\alpha-\gamma)\tau}}
\def\em{e^{-i\dm\tau}}
\def\emb{e^{+i\dm\tau}}

\def\etapp{\bar\eta_{+-}}

\def\phipp{\phi_{+-}}

\def\P{{\cal P}}
\def\PP{\bar{\cal P}}
\def\Q{{\cal Q}}
\def\Dt{\Delta\tau}
\def\O{{\cal O}}
\def\M{{\cal M}}
\def\ra{\rightarrow}
\def\Re{{\rm Re}}
\def\Im{{\rm Im}}
\def\qmg{{(\diamondsuit)}}
\def\rrho{{\hat{\rho}}}
\def\cpt{{\it CPT\/}}
\def\cptc{\cpt\ conservation}
\def\cptv{\cpt\ violation}
\def\cp{{\it CP\/}}
\def\cpv{\cp\ violation}

\def\phif{$\phi\,$ factory}
\def\phifs{$\phi\,$ factories}
\def\qm{quantum mechanics}

\REF\leeyang{T. D. Lee and C. N. Yang, \sl
Phys. Rev. \bf 98,  \rm 1501 (1955).}
\REF\cronin{J. H. Christenson, J. W. Cronin, V. L. Fitch and R. Turlay,
\sl Phys. Rev. Lett. \bf 13, \rm 138 (1964).}
\REF\cplear{CPLEAR collaboration, R. Adler et al., \sl Phys. Lett.
{\bf B286}, \rm 180 (1992).}
\REF\pecceione{C. D. Buchanan, R. Cousin, C. O. Dib, R. D. Peccei and J.
Quackenbush, \sl Phys. Rev. {\bf D45}, \rm 4088 (1992).}
\REF\frascati{{\it The DA$\Phi$NE PHYSICS HANDBOOK},  L.
Maiani, G. Pancheri and N. Paver,  eds. (INFN, Frascati, 1992). }
\REF\cptthm{R. F. Streater and A. S.
Wightmann, \sl PCT, Spin and Statistics, and
         All That.  \rm (Benjamin, New York, 1964).}
\REF\hawk{S. W. Hawking,  \sl
Phys. Rev. \bf D 14, \rm 2460 (1975), \sl Commun.
Math. Phys. \bf 87, \rm 395 (1982).}
\REF\dpage{D. N. Page, \sl Gen. Rel. Grav. \bf 14, \rm (1982);
L. Alvarez-Gaum\'e and C. Gomez, \sl Commun. Math. Phys. \bf 89,
 \rm 235 (1983);
 R. M. Wald,  \sl  General Relativity. \rm (Univ. of Chicago
Press, Chicago, 1984).}
\REF\pdb{Particle Data Group, K. Hikasa et al., \sl
Review of Particle Properties,  Phys. Rev. {\bf D45}, \rm
No. 11-II (1992). }
\REF\bps{T. Banks, M. E. Peskin and L. Susskind, \sl
Nucl. Phys. \bf B244, \rm 125 (1984).}
\REF\Eberhard{P.H. Eberhard, CERN Report No. CERN 72-1,
(1972),~(unpublished).}
\REF\Carithers{W.C. Carithers, J.H. Christenson, P.H. Eberhard, D.R.
Nygren, T. Modis, T.P. Pun, E.L. Schwartz and H. Sticker, \sl Phys.
Rev. {\bf D14}, \rm 290 (1976).}
\REF\ellishns{J. Ellis, J. S. Hagelin, D. V. Nanopoulos and M. Srednicki,
\sl Nucl. Phys.  \bf B241, \rm 381 (1984).}
\REF\zeil{A. Zeilinger, M. A. Horne, and C. G. Shull, in \sl Proceedings
    of the International Symposium on the Foundations of Quantum
           Mechanics, \rm  S. Kamefuchi, \etal, eds. (Physical Society
             of Japan, Tokyo, 1984).}
\REF\ellis{J. Ellis , N. E. Mavromatos and D. V. Nanopoulos, \sl
Phys. Lett.
\bf B293, \rm 142 (1992).}
\REF\ellistwo{J. Ellis, N. E. Mavromatos, and D. V. Nanopoulos,
 CERN-TH.6755/92 (1992).}
\REF\heidi{C. Geweniger, \etal., \sl
 Phys. Lett. \bf 48B, \rm 483  (1974).}
\REF\heidii{C. Geweniger, \etal,
 \sl Phys. Lett. \bf 48B, \rm 487 (1974).}
\REF\heidiii{S. Gjesdal, \etal,  \sl Phys. Lett. \bf 52B, \rm
113 (1974).}
\REF\Maianirev{L. Maiani, in the DA$\Phi$NE Handbook, ref. \frascati,
         vol. I.}
\REF\pecceiall{C.O. Dib and R.D. Peccei, \sl Phys.Rev. {\bf D 46},
\rm 2265 (1992).}
\REF\CERN{G. D. Barr, \etal, \sl Phys. Lett. \bf B317, \rm 233
               (1993).}
\REF\Winstein{L. K. Gibbons, \etal, \sl Phys. Rev. Lett. \bf 70,
            \rm 1203 (1993).}
\REF\dunietz{I. Dunietz, J. Hauser and J. Rosner, \sl Phys. Rev. \bf
D35, \rm 2166 (1987).}


\chapter{Introduction}

  The propagation and decays of
$\ko$ and $\kob$ mesons has provided elementary particle physicists
with the most fruitful system for probing the fundamental discrete
symmetries of Nature.  The paradoxes of this system led Lee and Yang
to postulate the violation of parity in the weak
interactions.\refmark\leeyang  More detailed exploration
led to the discovery of \cpv,\refmark\cronin and this system remains
the only place in which $CP$ violation  and $T$ violation have been
observed.
New experiments at
CPLEAR
 are now strengthening the precision of our knowledge of neutral
kaon physics,\refmark\cplear
 and beautiful new experiments on the discrete symmetries are
planned for future \phifs.\refmark{\pecceiall,\frascati}

  One of the goals of current and planned experiments on the
  $\ko$--$\kob$ system is to search for the violation of \cpt.
To interpret these experiments, one should have some idea of the
source of \cptv.
   Within the context of local quantum field theory, \cptc\ is
a theorem.\refmark\cptthm   Thus, theories of \cptv\ must necessarily
step outside the standard assumption that particle physics is
governed by a local quantum field theory.  In the 1960's, one could
plausibly build a theory of \cptv\ in the context of a nonlocal
model of the strong interactions.  However, with the triumph of the
standard model, that route is no longer open.

    On the other hand, developments in the quantum theory of gravity
have opened another possible avenue to theories of \cptv.  Building
from his results on the spectrum of radiation from black holes,
Hawking has proposed that the generalization of quantum mechanics
which encompasses gravity allows the evolution of pure states into
mixed states.\refmark\hawk   Page then showed that any such
dynamics also leads to conflict with \cptc.\refmark\dpage  These
ideas raised the interesting possibility that one could find observable
\cptv\ due to a mechanism that
lies not only beyond a local quantum description but also
beyond quantum mechanics altogether.  The notion that gravitation
effects beyond quantum mechanics can affect elementary particle
physics is controversial.  For example, in theories which allow the
evolution of pure states into mixed states, there is a serious
conflict between energy-momentum conservation and
locality.\refmark\bps  In this paper, however,  we will use a formalism
which avoids the issue of locality.  Our main goal will be to discuss
how the issue of the  violation of quantum mechanics can be addressed
experimentally.

   To our knowledge, the first attempt at a phenomenological analysis
of this type of quantum mechanics violation was made by
 P.H. Eberhard in the early
70s',\refmark\Eberhard as an attempt to probe the postulates of
quantum field theory. Eberhard suggested various tests of the
existence of a unitary S-matrix. His ideas motivated an experimental
test of quantum mechanics in the \kkbar\ system by Carithers
\etal.\refmark\Carithers
   The suggestion that Hawking's idea could be tested experimentally
in the \kkbar\ system was put forward independently
by Ellis, Hagelin, Nanopoulos,
and Srednicki (EHNS).\refmark\ellishns   EHNS set up an evolution
equation for the \kkbar\ system in the space
 of density matrices.  Their equation contains three new
 \cpt\ violating parameters $\alpha$, $\beta$, and $\gamma$. These
parameters have the dimensions of mass and might be expected to be
of order  $m_K^2/m_{\rm Pl} \sim 10^{-19}$ GeV.

  These authors were attracted to the \kkbar\ system by the fact that
it contains phenomena which depend on quantum coherence over
macroscopic distances. Though there are other phenomena that depend
on macroscopic quantum correlations, for example, the persistence of
superfluid and superconducting currents,  the \kkbar\ system gives a
controlled setting, involving only one particle, in which precision
experiments can reveal small deviations from the predictions of
quantum mechanics.  A second such system, also discussed by EHNS,
is found in macroscopic neutron interferometry.
Here, experiments of Zeilinger, Horne, and Shull have constrained a
similar dimensionful parameter of quantum mechanics violation to a
level  less than $0.8 \times 10^{-25}$ GeV,\refmark\zeil under the
assumption that quantum mechanics violation can randomize the neutron
spin.

Recently, Ellis, Mavromatos, and Nanopoulos
(EMN)\refmark{\ellis,\ellistwo}
reconsidered the analysis of  EHNS for the \kkbar\ system
and presented experimentally allowed
regions for the parameters $\alpha$, $\beta$, and $\gamma$
which were consistent with nonzero values of the magnitude of
the earlier estimates. Their analysis
suggested that this new source of \cptv\ might fully account for
the observed \cpv\ in the \kkbar\ system.
EMN also presented a microscopic theory of $\alpha$, $\beta$, and
$\gamma$, based on string theory, which gives values of the size of the
estimate above.
While we do not accept their detailed microscopic arguments, we
believe that their suggestion that quantum mechanics violation  might
be observable in present or planned experiments is an exciting  one
which deserves further consideration.  This is especially true
because the analysis of EMN was incomplete, as these authors
themselves pointed out, in that it did not fully take into account
constraints from the time evolution of the \kkbar\ state.

It is the purpose of this  paper to analyze the dynamics of the
 \kkbar \ system taking into account the possibility of  \cptv\ from
mechanisms both within and outside quantum mechanics.  We will
use this framework to discuss the implications of
 past, present, and future experiments on
the \kkbar\ system.  Our formulae will include both the conventional
\cptv\ within quantum mechanics and the \cptv\ outside quantum
mechanics of EHNS.  Our goal is to explain how to disentangle these
two possible sources of \cptv\ from one another, and how to
distinguish both from conventional \cpv.

    Our discussion will proceed as follows:  In Section 2, we will
review briefly the conventional parametrization of \cp\ and \cptv\
within quantum mechanics by the parameters $\ek$ and $\dk$.  We will
then rewrite this discussion in the language of density matrices.
In Section 3, we will review the formalism of EHNS and generalize
the density matrix equations to include their parameters
for quantum mechanics violation,
$\alpha$, $\beta$, and $\gamma$.

In Section 4, we will work out expressions for the
observable quantities in experiments on single neutral kaons
in terms of these parameters.
We will show that the parameter $\alpha$ is not significantly
constrained by these experiments, but
 that $\beta$ and $\gamma$  are tightly constrained by current data.
 Using the classic results of the  Carithers \etal\refmark\Carithers and
CERN-Heidelberg experiments\refmark{\heidi-\heidiii}
 and recent results from
CPLEAR,\refmark\cplear we determine the values $\beta
\,=\,(0.32\pm 0.29)\times
  10^{-18}$ GeV,
$\gamma\,= (-0.2 \pm 2.2)\times 10^{-21}$ GeV.
These bounds are
much stronger than those of EMN and lead to the simple
but important conclusion that the \cpv\ observed in the \kkbar\
system is dominantly quantum mechanical in nature and of \cpt\
conserving origin.   The analysis of Section 4  includes provision
for \cptv\ from within quantum mechanics, but it ignores the
possibility of \cptv\ in decay matrix elements.  In Section 5, we
introduce this possible additional complication and explain how it
potentially weakens our results.

 In Sections 6--8, we will discuss precision tests for \cptv\
and violation of quantum mechanics at the future \phifs.  A
\phif\ produces $K$ meson pairs in an antisymmetric pure state.
In ref. \pecceione, Peccei and collaborators showed that, within
the context of quantum mechanical \cp\ and \cptv, the full
time dependence of the decays into identical states, for example, to
$\{\pi^+\pi^-\,,\,\pi^+\pi^-\}$,  is independent of the \cp\ violation
parameters
and so provides an accurate measurement of the mass difference and
lifetimes of the kaon eigenstates.  When we allow for a departure
from quantum mechanics, this is no longer true.  We find an
additional time-dependent oscillation which, if observed experimentally,
would provide direct evidence of time evolution outside quantum
mechanics.  Section 6 provides the general formalism for describing
this effect. Section 7 gives formulae from which  $\alpha$, $\beta$,
and $\gamma$ can be constrained independently of one another
in specific \phif\ experiments.
In Section 8, we discuss
various processes in which the two kaons
decay asymmetrically. In particular, we consider the decay into
$\{\pi^+\pi^-\,,\,\pi^0\pi^0\}$, which provides a measurement of
Re $\ek'/\ek$, and we display corrections to the standard
formulae which appear when $\beta$ and $\gamma$ are nonzero.

\chapter{Quantum mechanics of a neutral kaon beam}

   In quantum mechanics, a kaon state evolves according to the action
of a Hamiltonian.  Even if the kaon decays, its evolution is correctly
described by an effective Hamiltonian which includes the natural width
of the state.

   One conventionally writes this effective Hamiltonian as
$$        H = M - {i\over 2} \Gamma,
\eqn\theham$$
where $M$ and $\Gamma$ are Hermitian $2\times 2$ matrices acting on
states in the basis $(\ket{K_0}, \ket{\bar K_0})$.  These states
are alternatively described using the basis of \cp\ eigenstates,
$$ \ket{K_1} = {1\over \sqrt{2}} \bigl(\ket{\ko} + \ket{\kob}\bigr) \ ;
 \qquad
\ket{K_2} = {1\over \sqrt{2}} \bigl(\ket{\ko} - \ket{\kob}\bigr)\ ,
\eqn\konetwodef$$
or the basis of eigenstates of $H$,  $\ket{K_S}$ and $\ket{K_L}$.

It is easiest to express predictions for the $\ko$--$\kob$ system by
trading the matrix elements of $H$ for parameters which express the
eigenvalues and eigenvectors of this matrix.  We will use the
conventions of Maiani,\refmark{\Maianirev} which express the
eigenstates of $H$ in terms of parameters $\ek_M$, which is odd under
\cp\ but even under \cpt, and $\dk$, which is odd under both \cp\ and
$\cpt$.  Explicitly,
$$  \eqalign{
 \ket{K_S} & = {N_S\over \sqrt{2}} \bigl( (1 + \ek_S)\ket{\ko}
                    + (1-\ek_S)\ket{\kob} \bigr) \crr
 \ket{K_L} & = {N_L\over \sqrt{2}} \bigl( (1 + \ek_L)\ket{\ko}
                    - (1-\ek_L)\ket{\kob} \bigr) ,  \cr}
\eqn\shortlongdef$$
where
$$   \ek_S = \ek_M + \dk \ , \qquad \ek_L = \ek_M - \dk,
\eqn\eksldef$$
and $N_S$, $N_L$ are real, positive normalization factors.  We write
the corresponding eigenvalues as
$$\eqalign{       \ms - {i\over 2}\gs &=  (\mb - \dm/2) - {i\over 2}
                          (\gb + \dg/2) \crr
     \ml - {i\over 2}\gl &=  (\mb + \dm/2) - {i\over 2}
                          (\gb - \dg/2) ,\cr}
\eqn\mandgdefs$$
so that $\dm$ and $\dg$ are positive.

   From this description of the kaon dynamics in terms of states, it
is easy to construct the description in terms of density matrices.
With the complex Hamiltonian \theham, the density matrix evolves
according to
$$
\rho_K(\tau)\, =\, e^{-iH\tau}\,\rk(0)\,e^{iH^{\dagger}\tau}  .
\eqn\timerho
$$
The eigenmodes of  this equation  are
$$
\eqalign{
\rl\,=&\,\ket{\kl}\bra{\kl}  \cr
\rs\,=&\,\ket{\ks}\bra{\ks}   \cr
\ri\,=&\, \ket{\ks}\bra{\kl}   \cr
\rib\,=&\,\ket{\kl}\bra{\ks} . \cr}
\eqn\eigenvectors
$$
A generic kaon beam can be decomposed into these modes.  If we write
the expansion coefficients at time $\tau = 0$ as  $A_L$, $A_S$,
$A_I$, $A_{\bar I}$, we find the following general form for the
 time evolution:
$$\rk(\tau)\,=\,A_L \,\rl\,
\egl\,+\,A_S\,\rs\,\egs\,+\,A_I\,\ri\,\egi\,\em\,+\,
A_{\bar I}\,\rib\,\egi\,\emb\, .
\eqn\rhotau$$

When $\rk$ describes a pure quantum state, $\rho_{L,S}$ and
$\rho_{I,{\bar I}}$ are as in \eigenvectors\ and the coefficient of
the interference term $A_I$ is correlated with $A_L$ and $A_S$.  If
we write the pure state as $\rk = \ket{K}\bra{K}$, with $\ket{K} =
a_L \ket{\kl} + a_S \ket{\ks}$, then $A_L = |a_L|^2$,
$A_S\,=\,|a_S|^2$ and $A_I\,=\, A_{\bar I}^*\, =\,a_L a_S^{\dagger}$.
When $\rk$ is a mixed state, we will find that these properties no
longer hold.

  In the $\ket{K_1}, \ket{K_2}$ basis, eq. \konetwodef, the four
eigenmodes of the density matrix take the form:
$$   \eqalign{
 \rl  =   \pmatrix{ |\ek_L|^2 & \ek_L \cr \ek^*_L & 1 \cr}
               & \qquad
    \rs  =   \pmatrix{ 1 & \ek_S^* \cr \ek_S & |\ek_S|^2 \cr}\crr
 \ri  =   \pmatrix{ \ek_L^* & 1 \cr \ek^*_L \ek_S& \ek_S \cr}
               & \qquad
 \rib =  \pmatrix{ \ek_L & \ek_S^* \ek_L\cr 1 & \ek_S^*\cr}.\cr}
\eqn\fourmodes$$
Here and henceforth, we normalize the eigenmodes $\rho_i$ so that
the largest element is 1; the normalization factors $N_S$, $N_L$
in \shortlongdef\ can be absorbed into the coefficients $A_i$
in \rhotau.

Any physical property of the kaon beam can be extracted from the
density matrix by tracing  with a suitable Hermitian operator.  To
extract the value of the observable ${\cal P}$, we write
$$  \VEV{{\cal P}} = \tr \bigl[ \rk  \O_\P \bigr].
\eqn\generalVEV$$
This expression for expectation values will remain true in the
generalization of quantum mechanics described in Section 3.
In the remainder of this section, we will write the operators
$\O_{\cal P}$ associated with the most important observables of
the  neutral kaon system.

 In principle, \cp\ and \cptv\ can show up not only in the neutral
kaon mass matrix but also in the kaon decay amplitudes.  There is already
some experimental evidence for a nonzero value of the parameter
$\ek'$ which characterizes direct \cpv\ in decays to
two pions.\refmark{\CERN,\Winstein}  However, the \cpv\ associated
with mass mixing is much more important.  In models such as the
superweak model, in which  \cpv\ is the result of new physics at a
scale $M$ much greater than $m_W$, direct \cpv\  typically results from
higher-dimension local operators and so is suppressed by
extra powers of $(m_W/M)$.  Thus, the observation of direct \cpv\ is
evidence for an origin of \cpv\ at the weak interaction scale, for
example, by the Kobayashi-Maskawa mechanism.

 Although we know strictly that  \cptv\ cannot  be the result of a
local quantum field theory, we believe it is nevertheless reasonable
to use the dimensional analysis rules of local quantum field theory
to restrict  possible sources  of \cptv.
This is certainly true in a model such as that of
ref. \ellistwo, in which violation of quantum mechanics arises from
applying an unusual averaging procedure to quantum mechanical
amplitudes.
When we consider models of quantum  mechanics violation, the  scale
$M$ which suppresses multipoint amplitudes must be very large.
 The appearance of \cptv\ within quantum mechanics will signal a
breakdown of local field theory, and this must take place at very
short distances compared to those probed in the LEP experiments at
the $Z^0$.  The \cptv\ outside quantum mechanics that we will
describe in the next section  has the Planck scale as its
characteristic mass scale.  Combining these ideas, we expect that
\cptv\ is associated with processes with the minimum number of
external particles.  Thus, in our main analysis, we will include
\cptv\ only in the time development of the neutral kaon state, and we
will ignore possible \cpt-violating contributions to the decay
vertices.

  Despite this argument, many phenomenological
 analyses  include the possibility of \cpt-violating decay amplitudes.
In this case, the experimental effects of these \cpt-violating terms
cannot be unambiguously disentangled from those of quantum mechanics
violation.  In the standard discussion without quantum mechanics
violation,
there is a similar difficulty in disentangling \cptv\ in the decay
amplitudes from that in the $K_L$--$K_S$ mass matrix.  In that case,
one can at least
present constraints
 on combinations of \cpt-violating parameters.  Then
one can
argue that either the individual parameters obey similar bounds
or there are unnatural large cancellations.\refmark\Maianirev
We will present the generalization of these constraints to the
theory with quantum mechanics violation in Section 5.

  By the same argument as that   which eliminates \cpt-violating
decay amplitudes,
   we will ignore the possibility
that these new processes violate the $\Delta S = \Delta Q$ rule for
leptonic $K$ decays.  In his review article, ref. \Maianirev,
Maiani has demonstrated that conventional contributions to
$\Delta S \neq \Delta Q$ amplitudes are also negligible.  Thus, we
will ignore
$\Delta S \neq \Delta Q$ effects throughout this paper.

  Within this set of assumptions, we now construct the operators
$\O_\P$ associated with the semileptonic and 2-pion decays of the
neutral kaons.  If we ignore violations of the $\Delta S = \Delta Q$
rule, only the $\ket{K_0}$ state can decay to $\pi^-\ell^+\nu$.
Then, using the basis $(\ket{K_1},\ket{K_2})$ of eq. \konetwodef,
the decay rate to this final state corresponds to the operator
$$  \O_{\ell^+}   =  |a|^2 \ket{K_0} \bra{K_0} = {|a|^2\over 2}
                            \pmatrix{ 1 &  1 \cr 1 & 1\cr},
\eqn\Oellplus$$
Similarly, the decay rate to $\pi^+ \ell^- \bar\nu$ is given by
$$  \O_{\ell^-}   =  |a|^2 \ket{\bar K_0} \bra{\bar K_0} = {|a|^2\over 2}
                            \pmatrix{ 1 &  -1 \cr -1 & 1\cr}.
\eqn\Oellminus$$
         In our expressions for the decay rate of neutral kaons to
two pions, we will ignore the possibility of direct \cptv, but we must
include an allowance for direct \cpv.  Then the amplitudes for
the decay of $\ket{\ko}$ and $\ket{\kob}$ to $\pi^+\pi^-$ are
$$\eqalign{ \M(\ko \ra \pi^+ \pi^-) & = A_0 e^{i\delta_0} +
              {1\over \sqrt{2}}A_2   e^{i\delta_2} \cr
 \M(\kob \ra \pi^+ \pi^-) & = A_0^* e^{i\delta_0} +
              {1\over \sqrt{2}}A_2^*   e^{i\delta_2}. \cr}
\eqn\piMsinkkbar$$
We choose the Wu-Yang convention in which $A_0$ is real.  Then the
decay amplitudes from states of definite \cp\ are:
$$\eqalign{ \M(K_1 \ra \pi^+ \pi^-) & = \sqrt{2} A_0 e^{i\delta_0} +
                         \Re A_2 e^{i\delta_2} \cr
 \M(K_2 \ra \pi^+ \pi^-) & = i \Im A_2
                        e^{i\delta_2}. \cr}
\eqn\piMsincp$$
Then, the decay rate to $\pi^+\pi^-$ is given, in the basis
\konetwodef,
by the operator
$$   \O_{+-}  = \pmatrix{|\sqrt{2} A_0 + \Re A_2 e^{i\delta}|^2 &
    (\sqrt{2} A_0 + \Re A_2 e^{-i\delta})(i \Im A_2 e^{i\delta}) \cr
    (\sqrt{2} A_0 + \Re A_2 e^{i\delta})(-i \Im A_2 e^{-i\delta}) &
    |i \Im A_2 e^{i\delta}|^2 \cr} ,
\eqn\Opmdef$$
where $\delta = \delta_2 - \delta_0$.  One can check this result
by tracing with the density matrices for the eigenstates $K_L$ and
$K_S$, as given in eq. \fourmodes.  We find
$$ \eqalign{
   \tr \bigl[ \O_{+-} \rl\bigr]
    & =   \bigl| \ek_L(\sqrt{2} A_0+ \Re A_2
                    e^{i\delta}) +  i \Im A_2 e^{i\delta} \bigr|^2 \cr
   \tr \bigl[ \O_{+-} \rs\bigr]
   & =   \bigl| \sqrt{2}A_0 + \Re A_2
                    e^{i\delta}) +  \O(\ek_s) \bigr|^2,\cr}
 \eqn\chargedpidecs$$
from which we recover, to leading order in \cpv\ and $\Delta I = 1/2$
enhancement, the familiar result\refmark{\Maianirev}
$$  |\eta_{+-}|^2 =
 \biggl|{\M(K_L\ra \pi^+\pi^-)\over \M(K_S\ra \pi^+\pi^-)}\biggr|^2
    = \bigl| \ek_L + {1\over \sqrt{2}} {i \Im A_2\over A_0} e^{i\delta}
            \bigr|^2.
\eqn\etapmfind$$

   The analogous argument for the neutral pion decay leads to the
operator
$$   \O_{00}  = \pmatrix{|\sqrt{2} A_0 -2 \Re A_2 e^{i\delta}|^2 &
    (\sqrt{2} A_0 -2 \Re A_2 e^{-i\delta})(-2i \Im A_2 e^{i\delta}) \cr
    (\sqrt{2} A_0 -2 \Re A_2 e^{i\delta})(2i \Im A_2 e^{-i\delta}) &
    |-2i \Im A_2 e^{i\delta}|^2 \cr} .
\eqn\Ozzdef$$
 From this expression, in the context of purely quantum-mechanical
evolution, we find
$$  |\eta_{00}|^2 =
 \biggl|{\M(K_L\ra \pi^0\pi^0)\over \M(K_S\ra \pi^0\pi^0)}\biggr|^2
    = \bigl| \ek_L - \sqrt{2} {i \Im A_2\over A_0} e^{i\delta}
            \bigr|^2.
\eqn\etapmfind$$

The quantities $\eta_{+-}$ and $\eta_{00}$ are parametrized in terms
of \cp-violating observables $\ek$ and $\ek'$:
$$    \eta_{+-} = \ek + \ek'\ , \qquad
      \eta_{00} = \ek - 2 \ek'\ .
\eqn\epsepsprimedef$$
When we ignore all effects outside quantum mechanics and also
ignore the possibility of \cptv\ in decay
 amplitudes, we find\refmark{\Maianirev}
$$    \ek = \ek_L = \ek_M - \Delta \ , \qquad
          \ek' = {i\over \sqrt{2}}{ \Im A_2\over A_0} e^{i\delta} .
\eqn\epsilonkqu$$

{}From this review of the effects of purely quantum-mechanical evolution
on the \kkbar\ system, we now procede to the  case of propagation
outside quantum mechanics.

 \chapter{Kaon evolution outside quantum mechanics}

  We now wish to enlarge the framework of our kaon beam equations of
motion to allow the possibility that pure states can evolve into
mixed states.  In this section, we add terms to the density matrix
equation to allow this possibility.  Our discussion will
follow the arguments of
Ellis, Hagelin, Nanopoulos, and Srednicki (EHNS).\refmark{\ellishns}
We will extend their work in providing formulae for
time-dependent interference phenomena.

  In conventional quantum mechanics, the density matrix obeys the
evolution equation
$$   i {d\over d\tau} \rho =   \bigl[ H\,,\,\rho \bigr] .
\eqn\densitymateq$$
This equation guarantees the conservation of probability
$$    {d\over d \tau} \tr[\rho ] = 0 ,
\eqn\firstrhozero$$
and also a set of higher identities
$$    {d\over d\tau} \tr [ \rho^n ] = 0,
\eqn\nextrhozero$$
which imply that the purity of the state is not changed by
quantum mechanical evolution.

     In a two-state system such as the neutral kaon system, the
density matrix can be expanded
$$          \rho = \rho^0 {\bf 1} +  \rho^i \sigma^i,
\eqn\rhoexpandsig$$
where $i = 0,1,2,3$ and $\sigma^i$ is a Pauli sigma matrix.
  The statement of conservation of probability
$\tr[\rho] = 1$ becomes $\rho^0 = \half$, and the statement that
$\rho$ has positive eigenvalues becomes
$$            (\rho^0)^2 \geq   \sum_{i=1}^3 (\rho^i)^2
\eqn\rhobound$$
  If the Hamiltonian is expanded in the same
way, the equation of motion can be written
$$       {d\over d\tau} \rho =  2 \epsilon^{ijk} H^i \rho^j \sigma^k .
\eqn\secondev$$

   The quantum mechanical description of the neutral kaon system is
only slightly more complicated.  Here we work with the non-Hermitian
Hamiltonian
\theham\ which  includes a provision for the kaon states to decay.
The evolution equation \timerho\ leads to the equation
$$       {d\over d\tau} \rho =  2 \epsilon^{ijk} M^i \rho^j \sigma^k
   -  \Gamma^0 \rho -  \Gamma^i (\rho^0 \sigma^i + \rho^i {\bf 1}).
\eqn\thirdev$$
The value of $\rho^0 =  2\tr[\rho] $ will now decrease. However,
the inequality \rhobound \  must still be satisfied, and it will be
as long as the matrix $\Gamma$ has two positive eigenvalues.

  Hawking\refmark{\hawk}
   proposed that the modifications of quantum-mechanical evolution
due to quantum gravity effects could be described by writing
a more general linear
equation for the density matrix.  The most general term that we could
add to \thirdev\ is
$$ -h^{0j}\rho^j {\bf 1} - h^{j0} \sigma^j- h^{ij} \sigma^i \rho^j
\eqn\addtoH$$
  There are two natural restrictions on these terms. First, they
should be consistent with probability conservation.  Second, they
should not decrease the entropy of the density matrix; though pure
states can evolve into mixed states, mixed states should not evolve
into pure states.  The first of these requirements sets $h^{0j} = 0$.
   The second requirement eliminates any
$h^{j0}$ terms, since these would
order the completely random distribution with $\rho = \half{\bf 1} $,
and also requires that the remaining submatrix $h^{ij}$
be positive definite.\refmark{\ellishns}   This leads to the following
set of equations for the components of the density matrix:
$$ \eqalign{
   {d\over d\tau} \rho^0 &=
   -  \Gamma^0 \rho^0 -  \Gamma^i \rho^i \cr
   {d\over d\tau} \rho^i &=  2\ek^{ijk} M^j \rho^k - \Gamma^i \rho^0
                   - \Gamma^0 \rho^i- h^{ij} \rho^j ,  \cr}
\eqn\rhodiffeqs$$
where, from here on, $i,j,k = 1,2,3$ only.  Notice that the antisymmetric
part of
$h^{ij}$ can be absorbed into $M^j$, so that we may assume from
here on that $h$ is symmetric.

   EHNS simplify this formalism
by imposing one further assumption.  If the action of quantum gravity
is universal among flavors, the new term  cannot change
strangeness.  Alternatively, if the basis chosen by the quantum gravity
interactions is close to the basis of quark mass eigenstates, the
strangeness nonconservation will be proportional to the square of
the rotation angle between these two bases.  If this angle is of
order the Cabibbo angle, strangeness violation will be a higher-order
effect.  Under either assumption,  we may concentrate on the part of
$h$ which conserves strangeness.  Since strangeness is measured
by the operator $\O_S = -\sigma^1$ in the basis of \cp\ eigenstates,
the restricted $\delta H$ must satisfy
$$   \tr[ \sigma^1  h^{ij}\sigma^i \rho^j]    = 0 \ , \qquad
             {\rm \ie,}\   h^{1j} = 0 .
\eqn\nooneH$$
This leaves
$$    h = 2 \pmatrix{ 0 & 0 & 0 \cr  0 & \alpha & \beta \cr
                              0 & \beta & \gamma \cr},
\eqn\abcdef$$
where the EHNS parameters $\alpha$, $\beta$, $\gamma$ satisfy
$$   \alpha, \gamma > 0\ , \qquad   \alpha\gamma > \beta^2 .
\eqn\abcrest$$

  The equations \rhodiffeqs\ for the density matrix components are
linear equations which can be solved by diagonalizing a $4\times 4$
matrix.  Since \cpv\ is a small effect, of order $10^{-3}$ of $\dm$
and $\dg$,
a perturbative solution
is quite appropriate.  To obtain this solution, we first  rewrite
the equations \rhodiffeqs\ in the basis of matrix elements of
the density matrix:
$$    \rho = \pmatrix{\rho_1 & \rho \cr  \bar\rho & \rho_2\cr}
\eqn\densitymat$$
It is useful to first write the purely quantum-mechanical equation
\thirdev\ in this basis:
$$  {d\over d\tau} \pmatrix{ \rho_1\cr \rho_2\cr \rho\cr \bar \rho\cr}
 = \left[ - \bar\Gamma + \pmatrix{
 -\dg/2 & 0 & +i\ek_L^* d^* & -i\ek_L d \cr
 0 & \dg/2 & +i\ek_S d & -i \ek_S^* d^* \cr
 -i\ek_S^* d^*& -i\ek_L d & +i \dm  & 0 \cr
 +i\ek_S d& +i\ek_L^* d^* & 0 & -i \dm  \cr} \right]
\pmatrix{ \rho_1\cr \rho_2\cr \rho\cr \bar \rho\cr},
\eqn\pqm$$
with
$$         d = \dm + {i\over 2} \dg .
\eqn\defofd$$
It is straightforward to check that the parameters are chosen so that
eigenvectors of the matrix reproduce \fourmodes, with the correct
eigenvalues.  The perturbation \abcdef\ adds to the quantity in
brackets the matrix
$$  \pmatrix{ - \gamma&  \gamma& -i\beta & + i \beta \cr
       \gamma& - \gamma& +i\beta & - i \beta \cr
        +i\beta & - i \beta & -\alpha &  \alpha\cr
        -i\beta & +i \beta & \alpha &  -\alpha\cr }
\eqn\EHNSpertub$$

Treating this addition to the equations in perturbation theory, and
working to the order of the leading contribution to each matrix
element, we can work out the new eigenmodes and eigenvalues.  The
eigenmodes are:
$$\eqalign{
\rl &=\pmatrix{ |\ek_L|^2 + \gamma/\dg + 4\beta(\dm/\dg) \Im[\ek_L/d^*]
               - \beta^2/|d|^2
 & \phantom{xxx}  \ek_L + \beta/d   \phantom{xxx} \cr
 \ek_L^* +\beta/d^* &   1  \cr} \crr
\rs &=\pmatrix{  1 &
 \ek_S^* -\beta/d^* \cr \phantom{xxx}   \ek_S - \beta/d \phantom{xxx}&
|\ek_S|^2 - \gamma/\dg - 4\beta(\dm/\dg) \Im[\ek_S/d^*]
               - \beta^2/|d|^2   \cr}\crr
\ri &=\pmatrix{ \ek_L^*-\beta/d^*&1\cr
-i \alpha/2\dm&\ek_S+\beta/d \cr}  \crr
\rib &= \pmatrix{ \ek_L -\beta/d& i \alpha/2\dm \cr
1&\ek_S^*+\beta/d^* \cr}  \cr}\,\,.
\eqn\fourmodestwo$$

The corresponding eigenvalues are corrected  by the shifts
$$\matrix{
   \gl \ra \gl + \gamma \ , & \gs \ra \gs + \gamma\ , \cr
       \gb \ra \gb + \alpha \ , &
 \dm\ra         \dm\cdot (1-{1\over 2}(\beta/\dg)^2)\ .  \cr }
\eqn\eigenvshifts$$
The shifts of $\Gamma_L$, $\Gamma_S$, and $\dm$ can be absorbed by
redefinition of these parameters.  The shift of $\dm$ is of relative
size $10^{-6}$ and so is negligible in any event.  If we redefine
$\bar \Gamma$ to be the average of the new $\Gamma_S$ and $\Gamma_L$,
then the interference terms $\ri$ and $\rib$ fall off at the
rate
$$        \bar\Gamma  + (\alpha - \gamma) \ .
\eqn\newbarGamma$$
This correction is not relevant to current experiments unless $\alpha$
is as large as $10^{-2}\gb$;
 in that case $\alpha$ would be 10 times larger
than  the familiar \cp-violating parameters.  We will retain this
shift, for completeness, in our formulae below, but we will
ignore it in our analysis of the present experimental situation.

  To summarize,  under the influence of quantum mechanics violation
as described by the formalism of EHNS,
   the most general  initial density matrix evolves according
to
$$\eqalign{
\rk(\tau)\,=\,A_L \,\rl\,
\egl &\,+\,A_S\,\rs\,\egs \cr
+\,&A_I\,\ri\,\egic\,\em\,+\,
A_{\bar I}\,\rib\,\egic\,\emb\, ,\cr}
\eqn\rhotautoo$$
 where now the eigenmodes are given by \fourmodestwo.

    To understand the changes that have been induced in the evolution
of density matrices, it is useful to rewrite the eigenmodes slightly in
order to emphasize their similarity to \fourmodes.  It is very
convenient to introduce the split $\ek$ parameters:
$$  \ek_L^\pm = \ek_L \pm \beta/d \ ,
  \ek_S^\pm = \ek_S \pm \beta/d \ .
\eqn\eksplit$$
Then the expressions for the eigenmodes can be rewritten as follows:
$$\eqalign{
\rl &=\pmatrix{ |\ek_L^-|^2 + \gamma/\dg + 4(\beta/\dg)
\Im[\ek_L^- d/d^*]
 & \phantom{xxx}  \ek_L^+\phantom{xxx} \cr
 \ek_L^{+*} &   1  \cr} \crr
\rs &=\pmatrix{  1 &
 \ek_S^{-*} \cr \phantom{xxx}   \ek_S^-\phantom{xxx}&
|\ek_S^+|^2 - \gamma/\dg - 4(\beta/\dg)\Im[\ek_S^+ d/d^*]
              \cr}\crr
\ri &=\pmatrix{ \ek_L^{-*}&1\cr
-i \alpha/2\dm&\ek_S^+\cr}  \crr
\rib &= \pmatrix{ \ek_L^-& i \alpha/2\dm \cr
1&\ek_S^{+*} .\cr}  \cr}
\eqn\fourmodestwo$$
The eigenmodes are written here to sufficient accuracy for the analysis
in the rest of this paper.  The complete expressions for these eigenmodes
to second order in small parameters are given in the Appendix.

Except for their smallest matrix elements, of which only $\rho_{L1}$
and  $\rho_{S2}$ are important in practice, the expressions
\fourmodestwo\ take precisely the form of the density matrices of
pure states given in \fourmodes.  However, the various modes contain
differently shifted $\epsilon$ parameters.  We will see in the next
section that different neutral kaon experiments are sensitive to
different choices of these parameters, in such a way that the various
sources of \cptv\ can be distinguished.

\chapter{Experimental determination of $\beta$ and $\gamma$}

In this section, we will show how present experimental data constrains
the EHNS parameters.  We will show how to combine the very accurate
measurements of the $K^0$ parameters from the experiments of the
early 1970's with new data on time-dependent $K^0$ evolution from
the CPLEAR experiment.  This comparison will give stringent
constraints on $\beta$ and $\gamma$ which limit their effects to
be at most about 10\% of the observed \cpv\ in the neutral kaon system.
This rules out the possibility, suggested in ref. \ellistwo, that
a deviation from \qm\ of the nature considered here is the major
source of observed \cpv.  We will also recommend a parametrization of
the future, more accurate CPLEAR data which will facilitate comparison
of this data with the EHNS formalism.

   In this section, we will consider specifically the following two
observables of the neutral kaon system:
$$ \eqalign{
 R_{+-}(\tau)
  &= {N(K(\tau)\ra \pi^+\pi^-)\over  N(K(\tau = 0) \ra \pi^+\pi^-)} \crr
 \delta(\tau)
  &= {N(K(\tau)\ra \pi^-\ell^+ \nu) -
N(K(\tau)\ra \pi^+\ell^- \bar\nu)
\over N(K(\tau)\ra \pi^-\ell^+ \nu) +
N(K(\tau)\ra \pi^+\ell^- \bar\nu) } \cr}
\eqn\obsdefs$$
These quantities are given in terms of density matrices and the
decay operators defined in Section 2 by
$$
 R_{+-}(\tau) = {\tr \rk(\tau)\O_{+-}\over  \tr \rk(0)\O_{+-} } \ , \quad
 \delta(\tau)
  = {\tr \rk(\tau) \bigl(\O_{\ell^+} - \O_{\ell^-}\bigr) \over
\tr \rk(\tau) \bigl(\O_{\ell^+} + \O_{\ell^-}\bigr) }
\eqn\obsrhodef$$

In  our analysis, we may ignore the effects of $\Im A_2$ in $\O_{+-}$,
since $\ek'/\ek \sim 10^{-4}$.  Thus, the formulae we present for
the charged pion decay apply equally well to the neutral pion decay.
It is easy to restore the effect of $\ek'$ by evaluating \obsrhodef\
more exactly.

  In a $K^0$ beam which has evolved to large $\tau$, the quantities
$\delta$ and $R_{+-}$ reflect the pure $K_L$ eigenstate.  In this
case, we can evaluate \obsrhodef\ with the density matrix $\rl$ and
obtain the following results:
$$   \eqalign{
 \delta_L & =2 \Re\, \ek_L^+  \cr
  R_L & = \rho_{L1} =
 |\ek_L^-|^2 + \gamma/\dg + 4(\beta/\dg)
\Im[\ek_L^- d/d^*] \cr}
\eqn\obslongvals$$

Alternatively, we could consider the evolution of a pure $K^0$ or
$\bar K^0$ state created in a strong interaction process.  For
state which is initially pure $K^0$, the evolving density matrix is
given by \rhotautoo\ with $A_L = A_S = A_I = A_{\bar I} = \half$,
up to corrections of order $\ek,\alpha$. (For an initial $\bar K^0$,
reverse the sign of the interference terms.) Then the time-dependent
quantities \obsdefs\ are given by\foot{This formula includes
all corrections linear in $\alpha$.}
$$ \eqalign{
 \delta(\tau) & = {2\cos(\dm \tau)\egic + 2\Re\, \ek_S^- \egs +
                       2 \Re\, \ek_L^+ \egl\over \egs + \egl
                    } \crr
 R_{+-}(\tau) &=   \egs +  R_L\egl +
          2 |\bar \eta_{+-}| \cos(\dm  \tau +\phi_{+-})\egic \cr}
\eqn\timedefvals$$
with  $R_L$ as in \obslongvals, and
$$  |\bar \eta_{+-}|e^{i\phi_{+-}}=  \ek_L^-  \ .
\eqn\etaandphi$$

  Notice that the interference term in the time-dependent asymmetry
depends on $\ek_L^-$, while the $K_L$ lepton asymmetry depends on
$\ek_L^+$.  Thus, by comparing time-dependent and long-time
 measurements, we can find a
constraint on $\beta$ which is independent of other sources of
\cptv.  By comparing the value of $|\ek_L^-|^2$ to a determination
of $R_{L}$, and using this bound on $\beta$, we can also obtain a
 bound on $\gamma$.

   The difficulty in implementing this program is that experiments on
the time-dependent evolution of neutral kaon beams typically
report fits of their data to the conventional \cpt-conserving
theory, in which $\ek_L^-$, $\ek_L^+$, and the square root of
$\rho_{L1}$ are not distinguished.  Thus, it is important to
understand which particular parameters are actually constrained by
each given experiment.

  The value of $\delta_L$ poses no difficulty.  The current world
average\refmark\pdb gives
$$          \delta_L   = 2 \Re\, \ek_L^+
= (3.27 \pm 0.12)\times 10^{-3} \ .
\eqn\thedeltaL$$

  In the early 1970's, the CERN-Heidelberg collaboration carried out
beautiful studies of the time-dependence of $K_L$--$K_S$
 interference.\refmark{\heidi-\heidiii}  These studies confirmed that
the conventional, \cpt-conserving parametrization of the neutral
kaon system indeed gives a good description of its time-dependent
phenomena.  However, it is  very difficult to reconstruct
the constraints that these experiments put on more general models
of time-dependence.  The experiments involved kaon production from
a platinum target; in this situation, the proportion of $K^0$ versus
$\bar K^0$ initial states is not known {\it a priori} and is also
momentum-dependent.   The particle-antiparticle asymmetry must
be described by an
unknown function
$$        A(p) =
{N(p)-{\bar N}(p) \over N(p)+{\bar N}(p)}
\eqn\defofSp$$
 In ref. \heidii, data on neutral kaon decays
to $\pi^+\pi^-$ was binned in momentum and decay time and these
distributions were used to fit for the parameters $|\eta_{+-}|$,
$\phipp$, and $A(p)$ (in each bin) from the assumed relation
$$
R_{+-}(\tau)= \egs+ |\eta_{+-}|^2 \egl +
 2 A(p) |\eta_{+-}| \cos (\dm\tau + \phipp)\egi .
\eqn\assform$$
Unfortunately, the absolute magnitude of the interference term can be
absorbed into the parameters $A(p)$, so it is not possible to
determine the coefficients $R_L$ and $|\bar\eta_{+-}|$ separately.
Since the strongest constraint on $|\eta_{+-}|$    from this data set
comes from the long-time tail of the decay distribution, we will
consider the CERN-Heidelberg determination of $|\eta_{+-}|$ to be a
measurement of $R_L$:
 $$       \sqrt{R_L} = (2.30 \pm  0.035) \times 10^{-3} .
\eqn\rhofind$$
{}From Fig. 4 of ref. \heidii, it is clear that quantum mechanics
does correctly describe the interference region to few-percent
accuracy after adjustment of the $A(p)$, but we will not make use
that information
in the discussion below.

  Shortly afterward, the experimental group  of Carithers
\etal, working at the LBL Bevatron, used their data on $K\ra \pi\pi$
decays in a regenerated $K_L$ beam to directly measure the
magnitude of the interference term in neutral kaon
evolution.\refmark\Carithers    In
the notation of eq. \timedefvals, they obtained the result
$$  { |\bar\eta_{+-}|\over \sqrt{R_L} } = 0.972 \pm 0.021  \ .
\eqn\Carithres$$
  Very recently, the CPLEAR collaboration has published as its
first result a determination of $|\etapp|$ using hadronically
produced neutral kaons tagged by an accompanying charged
kaon.\refmark{\cplear}
Their determination is dominated by the interference region, and so
we may interpret their result as a measurement of $|\bar\eta_{+-}|$.
This gives
$$               |\bar\eta_{+-}| = (2.32 \pm 0.14)\times 10^{-3} \ .
\eqn\ekmagdeta$$
Averaging these measurements using the CERN-Heidelberg value for
$R_L$, we find
$$               |\bar\eta_{+-}| = (2.249 \pm 0.054)\times 10^{-3} \ .
\eqn\ekmagdet$$

  Finally, we see no difficulty in accepting the world average
of measurements of the interference phase $\phi_{+-}$ as a
determination of the phase of $\ek_L^-$:
$$               \phi_{+-} = (46.5 \pm 1.2)^\circ
\eqn\ekphasedet$$

  The comparison of these four numbers allows us to constrain $\beta$
and $\gamma$.  We can find $\beta$ from the relation
$$   \Re{2\beta\over d}  = \Re\, \ek_L^+ - \Re\, \ek_L^-  \ .
\eqn\constrbeta$$
To analyze this relation, we use the fact that $\beta$ is real and
$d$ is very well known:\refmark\pdb
$$    d = \dm + {i\over 2} \dg = \bigl((3.522\pm 0.016) + i
                   (3.682 \pm 0.008) \bigr) \times 10^{-15} {\rm GeV}
\eqn\valofd$$
It is convenient to parametrize
$$    d = |d| e^{i(\pi/2-\phi_{SW})}  \ ,
\eqn\superwdef$$
where
 $\phi_{SW} = (43.73 \pm 0.15)^\circ$ is the {\it
superweak phase} of Maiani's conventions.\refmark\Maianirev
  With this definition
$$    \beta = {|d|\over 2 \sin\phi_{SW} } \bigl( {\delta_L \over 2}
               - |\bar\eta_{+-}| \cos \phi_{+-} \bigr)
\eqn\betaeqn$$
This yields
$$    \beta =  (3.2 \pm 2.9) \times 10^{-19}\ {\rm GeV} \ .
\eqn\valofbeta$$
With this determination of $\beta$, we can evaluate $\gamma$ by
comparing $R_L$ to $\ek_L^-$:
$$   \gamma = \dg \bigl[  R_L   -
 |\bar\eta_{+-}|^2 - 4(\beta/\dg)
  |\bar\eta_{+-}|\sin( 2 \phi_{SW}-\phi_{+-})\bigr]
 \  .
\eqn\findgamma$$
which gives
$$  \gamma =  (- 0.2 \pm 2.2) \times 10^{-21} \ {\rm GeV}\ .
\eqn\valofgamma$$
These constraints are similar to those obtained by Ellis, Mavromatos,
and Nanopoulos\refmark{\ellis,\ellistwo}  in one-parameter fits for
$\beta$ and $\gamma$.  However, our constraints hold in the general
three-parameter EHNS model; they exclude the possibility that $\beta$
and $\gamma$ could give large individual
contributions which cancel in $R_L$.

      \FIG\constb{Constraints on the EHNS parameters $\beta$ and $\gamma$
            from the comparison of determinations of the $\ek$
             parameter from different observables of  the
             $K_L$--$K_S$
           system:  (a)  the systematics of expected discrepancies;
            (b) the   current experimental situation.}
   The geometry of these constraints is shown in Fig. \constb.
In Fig. \constb (a), we show the systematics of the various
values of $\ek$.   This figure should be compared to Fig. 1 of
ref. \ellis.
 The parameter $\ek_L^-$, considered as a vector
in the complex plane, is directly determined by the interference
measurement.  From the endpoint of this vector, $\ek_L^+$ is found
by making an excursion downward at 45$^\circ$ by a displacement
proportional to $\beta$.  Similarly, $R_L$ is found by
moving outward a distance proportional to $\gamma$, after correcting
for a $\beta$ effect.  In Fig. \constb (b), we show the constraints
on these excursions given by the experimental values \thedeltaL,
\rhofind,  \ekmagdet, \ekphasedet.  The various constraints overlap
in the $\ek$ plane  in such a way as to constraint $\beta$ and
$\gamma$ to contribute only a very small part of the \cpv\
phenomenology.

   It is important to note that the constraints on \cptv\ outside
of quantum mechanics that we have considered here are quite
independent of the possibility of \cptv\ within quantum mechanics.
  Such a source of
\cptv\ can be constrained, just as in the analysis without quantum
mechanics violation,\refmark\Maianirev
 by verifying the extend to which
 $\ek_L = \ek_L^- + \beta/d$ is parallel to $d$ in the complex
 plane.   We will give the precise argument in the next section.
 Since we have not used any  information
on the absolute
phase of $\ek_L$ in order to bound $\beta$ and $\gamma$,
the standard constraints on \cptv\ within quantum mechanics are
not significantly weakened when we allow for the presence of the
EHNS parameters.  In the future, the measurement of $\ek_S^\pm$
will allow a stronger constraint on this type of \cptv.

   To derive further constraints on the violation of quantum mechanics,
we strong\-ly recommend that time dependent decay distributions be
fit directly to the formulae \timedefvals, to determine the five
independent parameters $R_L$, $\delta_L$, $|\bar\eta_{+-}|$,
$\phi_{+-}$, $\alpha$.  This will make it possible to check, without
the ambiguities of our analysis here, whether the rapport among these
parameters definitively excludes the presence of new terms in the
density matrix evolution equation.

\chapter{Effects of \cptv\ in decay amplitudes}

    At the end of Section 2, we argued that, whether \cptv\  arises
from the breakdown of quantum mechanics or from the breakdown of
local quantum field theory within quantum mechanics, the effects
of direct \cptv\ in   decay amplitudes should be small compared to
 those in neutral kaon propagation. Nevertheless, much emphasis is
given in the literature on \cptv\ to disentangling effects of
\cptv\ in decay vertices from \cptv\ in the kaon mass matrix.
In this section, we will show that this separation can still be
made, and constraints on $\beta$ and $\gamma$ deduced, if one
allows for \cpt-violating additions to the kaon decay vertices.
However, this analysis will require additional assumptions which,
though not unreasonable, are not airtight.

To begin the analysis, we need the more complicated forms of the
decay operators $\O_{\cal P}$ which allow for \cptv\ in the
decay amplitudes.
  The parametrization  given in
Maiani's review article\refmark\Maianirev leads to the following
expressions, which replace \Oellplus, \Oellminus, \Opmdef, and
\Ozzdef:  for the leptonic decay amplitudes,
$$  \O_{\ell^+}   =  {|a+b|^2\over 2}
                            \pmatrix{ 1 &  1 \cr 1 & 1\cr}\ ,\qquad
 \O_{\ell^-}   =  {|a-b|^2\over 2}
                            \pmatrix{ 1 &  -1 \cr -1 & 1\cr} \ ;
\eqn\Oellminustwo$$
for the $\pi\pi$ decay amplitudes,
$$  \O_{+-}   =  |X_{+-}|^2
          \pmatrix{ 1 &  Y_{+-} \cr Y^*_{+-} & |Y_{+-}|^2\cr}\ ,\qquad
 \O_{00}   =  |X_{00}|^2
   \pmatrix{ 1 &  Y_{00} \cr Y_{00}^* & |Y_{00}|^2\cr},
\eqn\Opptwo$$
where
$$   X = \VEV{\pi\pi \Bigm| K_1 } \ ,\qquad
Y =   { \VEV{\pi\pi \bigm| K_2} \over\VEV{\pi\pi \bigm| K_1}} .
\eqn\Oppdefs$$
More explicitly,
$$ \eqalign{
 Y_{+-} &= \bigl( {\Re B_0\over  A_0}
                         \bigr) + \ek' \cr
 Y_{00} &= \bigl( {\Re B_0\over  A_0}
                         \bigr) -2 \ek' \cr}
\eqn\Ydefs$$
The quantities $\Re(B_0/A_0)$ and $\Re(b/a)$  parametrize
\cpt-violating decay amplitudes;
$\epsilon'$ has the \cpt\ conserving value \epsilonkqu\ shifted by the amount
$  e^{i\delta}(\Re B_0 /A_0 - \Re B_2/A_2)/{\sqrt 2} $
which accounts for a different degree of \cpt\ violation in the isospin
$I=0$ and $I=2$ pion-decay channels.\refmark\Maianirev
 With this generalization, the value of $\delta_L$ becomes
$$   \delta_L = 2 \Re \bigl[ \ek_L^+ + {b\over a} \bigr]
\eqn\newdeltaL$$
and the parameters $R_L$, $|\bar\eta_{+-}|$, and $\phi_{+-}$
in \timedefvals\ are shifted to
$$\eqalign{
R_L &= {\gamma/\dg}+|\bar\eta_{+-}|^2
+4(\beta/ \dg) \Im\bigl[ \bar\eta_{+-} d/d^*- Y_{+-}   \bigr]\cr
|\bar\eta_{+-}| e^{i\phi_{+-}} &=    \ek_L^-  + Y_{+-}\ . \cr}
\eqn\changethetimedefs$$

 When we include these corrections into the relation between $\ek_L^-$
and $\ek_L^+$, we find, instead of \betaeqn,
 the following expression for $\beta$:
$$    \beta  + {|d|\over 2\sin\phi_{SW}}\Re\bigl(
         {b\over a} - {B_0\over A_0} \bigr)
= {|d|\over 2 \sin\phi_{SW} } \bigl( {\delta_L \over 2}
               - |\bar\eta_{+-}| \cos \phi_{+-} \bigr) \ .
\eqn\betaeqntwo$$
The relation for $\gamma$ remains
$$   \gamma   = \dg \bigl[  R_L -
 |\bar\eta_{+-}|^2 - 4(\beta/\dg)
  |\bar\eta_{+-}|\sin( 2 \phi_{SW}-\phi_{+-})\bigr]
\eqn\gammatwo$$
if we ignore a very small correction proportional to $\beta\cdot
\Im \, Y_{+-} = \beta\cdot\Im\, \ek'$.
 Thus, our previous
constraints on $\beta$ and $\gamma$ now appear as constraints on
combinations of \cpt-violating parameters.

   For completeness, we should add one further constraint on a
combination of parameters characterizing  \cptv\ within and outside
of quantum mechanics.  This constraint, which is reviewed in
ref. \Maianirev, uses unitarity and the dominance of the isospin
0 $\pi\pi$ decay channel to determine the phases of the
\cpt-conserving and \cpt-violating mixing parameters.  If we
denote these components of the kaon mixing parameters as $\ek_M$ and
$\Delta$, as in eq. \eksldef,  then this constraint implies that
the phase of $\ek_M$ is $\phi_{SW}$ of eq. \superwdef, and the
phase of the combination
$$            \Delta - {\Re B_0\over A_0}
\eqn\deltacombo$$
is $\phi_{SW} \pm \pi/2$.

   To apply this relation, rewrite the second of eqs.
\changethetimedefs\ as:
$$    |\bar\eta_{+-}|e^{i\phi_{+-}}= \ek_M - \bigl(\Delta - {\Re B_0\over
              A_0} \bigr) - {\beta\over d} \ .
\eqn\anglerel$$
Notice that, because $d$ points at $45^\circ$, the last term has a
phase of ($-45^\circ$), so that it is also perpendicular to
$\ek_M$.  On the other hand, the experimental value of $\phi_{+-}$
 is very close to that of $\phi_{SW}$, though there is a small
discrepancy:
$$          \phi_{+-} - \phi_{SW} = (2.8 \pm 1.2)^\circ  \ .
\eqn\phidiscrep$$
Thus, by comparing the components of the right and left hand sides
of \anglerel\ perperdicular to $\ek_M$, we obtain a third constraint
on the parameters of \cptv.

This additional constraint reads:
$$   \beta \pm |d| \,\biggl|\Delta - {\Re B_0\over A_0} \biggr|  =
               (- 5.6 \pm 2.5)  \times 10^{-19} \ {\rm GeV} \ .
\eqn\thirdconst$$

%
This must be combined with the results of eqs. \betaeqntwo\ and
\gammatwo:
$$  \eqalign{
    \beta  + {|d|\over 2\sin\phi_{SW}}\Re\bigl(
         {b\over a} - {B_0\over A_0} \bigr)
&= (3.2\pm 2.9) \times 10^{-19} \ {\rm GeV} \cr
   \gamma -  2 |\bar \eta_{+-} d| \Re\bigl({b\over a}-{B_0\over A_0}
  \bigr)    &= (-0.2 \pm 2.2) \times
             10^{-21}\ {\rm GeV} \ . \cr}
\eqn\gammathree$$
In the last relation, we have used the approximation $\phi_{SW}
\approx \phi_{+-}$.
These three equations provide three constraints on four parameters
and so cannot disprove the existence of \cptv.  However, they imply
that, unless there are unnatural cancellations among these parameters,
the magnitude of \cptv\ should  be at most about a tenth that of
\cpv.

\chapter{Tests of \qm\ at a \phif: formalism}

   A high-luminosity \phif\ has been recognized as a facility which
gives particularly incisive tests of \cptv.\refmark\frascati
In models in which \cptv\ arises within quantum mechanics, Peccei
and collaborators have shown how, by studying the full set of
possible time-dependent asymmetries observable at a \phif, one may
disentangle \cpt-violating terms in the neutral kaon mass matrix
from \cpt-violating contributions to kaon decay
amplitudes.\refmark{\pecceione, \pecceiall}    This
analysis is made possible by the very simple time-dependence
predicted for the
$K_0$--$\bar K_0$ state which evolves from the decay of the $\phi$.
Since the $\phi$ has spin~1, its decay to two spinless bosons
produces an antisymmetric spatial wavefunction. This means that,
when the $\phi$ decays to two neutral kaons, those particles must
remain in opposite mass eigenstates until one decays.  This strong
constraint from quantum-mechanical coherence governs the whole
phenomenology of \phif\ measurements.

   Because of the importance of quantum-mechanical coherence in
$\phi$ decays, a \phif\ is also an ideal place to search for
terms in the kaon evolution which violate quantum coherence.  In
this section, we will explain how the standard formulae for
kaon correlations in $\phi$ decay are changed by the introduction
of quantum mechanics violation according to the model of EHNS,
and we will point out particularly incisive measurements for
determining or constraining the EHNS parameters.

  The key to our analysis will be the form of the density matrix
for the two-particle system which results from $\phi$ decay.  We will
first review the form of this density matrix in the case in which
\cpt\ is violated only by corrections within quantum mechanics.  Then
we will show how this result is generalized in the EHNS model.

 In
the following discussion, we will assume that the $\phi$ resonance
is pure spin~1, with no quantum mechanics violation in its decay
amplitudes.  Throughout our analysis, we will ignore background
processes which produce $K_0 \bar K_0$ in a spin~0 combination, and
also effects of finite detector size.  The influence of these
 effects in conventional \phif\ analyses are described in
ref. \frascati.

  A spin 1 $\phi$ meson decays to an antisymmetric state of
two kaons.  If the kaons are neutral, we can describe the resulting
wavefunction, in the basis of \cp\ eigenstates $\ket{K_1}$, $\ket{K_2}$,
as
$$
\phi\rightarrow{1\over {\sqrt 2}}
\Bigl( |K_1,p>\otimes|K_2,-p>-  |K_2,p>\otimes|K_1,-p> \Bigr)\  .
\eqn\phistate
$$
The two-kaon
density matrix resulting from this decay
is a $4 \times 4$ matrix.   We can express this matrix concisely by
introducing the set of $2\times 2$ matrices appropriate to the
$\ket{K_1}$, $\ket{K_2}$  basis:
$$
\rrho_1=\pmatrix{1&0\cr0&0\cr}\ , \quad
\rrho_2=\pmatrix{0&0\cr0&1\cr}\ , \quad
\rrho_+=\pmatrix{0&1\cr0&0\cr}\ , \quad
\rrho_-=\pmatrix{0&0\cr1&0\cr}\ .
\eqn\rodef$$
Then the state \phistate\ corresponds to the density matrix
$$ P = \half\Bigl(\rrho_1\otimes \rrho_2 + \rrho_2\otimes \rrho_1
-\rrho_+\otimes\rrho_- -\rrho_-\otimes\rrho_+ \Bigr) \ .
\eqn\phidensity$$

  To work with the density matrix \phidensity, we must express
its components $\rho_i$ in terms of eigenstates of the single-particle
density matrix evolution equation.  Then we can assign
each eigenstate its
natural time-dependence. This procedure accounts correctly the full
time-dependence of correlations.

 To see how this works, we will first express $P$ in terms of
eigenmodes in the quantum-mechanical case.  It is straightforward to
express the components \rodef\ in terms of the four matrices
\fourmodes\ and then to express the combination \phidensity \
in terms of these elements.  One finds, to no great surprise, an
antisymmetric combination of $\ket{K_L}$ and $\ket{K_S}$.  Here and
for the rest of this section, we will denote results derived
assuming quantum mechanical evolution (but not \cpt\ symmetry) with
a superscript diamond.  Thus,
$$ P^\qmg = {1 + 2 \Re(\ek_S \ek_L)\over 2}
\Bigl(\rs\otimes \rl + \rl\otimes \rs
-\ri\otimes\rib -\rib\otimes\ri\Bigr) \ .
\eqn\phidensity$$
The prefactor, which we have written to order $\ek^2$, corrects for
the fact that
$\ket{K_L}$ and $\ket{K_S}$ are not orthogonal.  Supplying the
proper time-dependence, we  find the density matrix for
processes in which the first kaon decays at proper time $\tau_1$ and
the second at proper time $\tau_2$.  We find
$$ \eqalign{
P^\qmg (&\tau_1,\tau_2) ={1 + 2\Re(\ek_S \ek_L)\over 2}
\Bigl(\rs\otimes\rl e^{-\gs\tau_1} e^{-\gl\tau_2} + \rl\otimes \rs
         e^{-\gl\tau_1}e^{-\gs\tau_2}  \cr  &\hskip 0.05in
-\ri\otimes\rib e^{-i\dm (\tau_1-\tau_2)}e^{-\gb(\tau_1+\tau_2)}
 -\rib\otimes\ri
 e^{+i\dm (\tau_1-\tau_2)}e^{-\gb(\tau_1+\tau_2)}\Bigr) \ .\cr}
\eqn\phidensitytime$$

  It is equally straightforward to perform this computation when
quantum mechanics violation is included.  One must first work out
expressions for the components \rodef\ in terms of the eigenmodes
\fourmodestwo.  To first order in small quantities, we find
$$\eqalign{
\rrho_1 &=  \rs +(\gamma/\dg)\rl   - \ek_S^{-*} \ri - \ek_S^-\rib \cr
\rrho_2 &= -(\gamma/\dg)\rs +  \rl -\ek_L^+ \ri - \ek_L^{+*}\rib \cr
\rrho_+ & = - \ek_L^{-*} \rs - \ek_S^+ \rl + \ri + i(\alpha/2\dm)\rib\cr
\rrho_- & = - \ek_L^{-} \rs - \ek_S^{+*}\rl-i(\alpha/2\dm)\ri
 +  \rib \ . \cr}
\eqn\rhopartsexpand$$
Inserting these expressions into \phidensity, we find
$$ \eqalign{
P = \half \Bigl[ \rs \otimes& \rl + \rl \otimes \rs - \ri\otimes \rib
           -\rib \otimes \ri \cr
     &   -2 {\beta\over d}(\rs\otimes \ri + \ri \otimes \rs)
        -2 {\beta\over d^*}(\rs\otimes \rib + \rib \otimes \rs) \cr
     &   +2 {\beta\over d^*}(\rl\otimes \ri + \ri \otimes \rl)
        +2 {\beta\over d}(\rl\otimes \rib + \rib \otimes \rl) \cr
& + i{\alpha\over \dm}\bigl( \ri\otimes \ri - \rib\otimes \rib\bigr)
+ {2\gamma\over \dg} \bigl(\rl\otimes \rl - \rs\otimes \rs \bigr)
 \Bigr]\ .  \cr}
\eqn\Pform$$
In eq. \Pform, each coefficient is given to first order in small
quantities.   We will carry out our analysis to this order.  However,
 if some of the EHNS parameters are  more   severely constrained than
others, one might need the expressions of second order in the
less constrained parameters to determine the correct limits for the
more constrained ones.  For example, we saw in  Section 4 that the
terms quadratic in $\beta$ affect the limits on $\gamma$.  In case a
similar situation arises in \phif\ experiments, we have given in the
Appendix the
 complete formula for $P$ correct to second order
in small quantities.

 When we study time-dependent correlations, each term of \Pform\ will
lead to a characteristic exponential behavior.  The first line of
\Pform\ is identical (within the approximation given) to \phidensity\
and thus contains only time-dependences allowed within quantum mechanics.
The next three lines of \Pform, however, contain completely new
structures.  The second line of \Pform\ leads to terms in the
decay distribution which behave as
$$            \cos(\dm \tau_1 - \phi) e^{-(\gb+\alpha-\gamma) \tau_1}
                      e^{-\gs\tau_2} \ ,
\eqn\newtimeone$$
and, in the same way, with 1 and 2 interchanged.  The third line
leads to  similar expressions with the decay rate $\gl$. The first
term in the fourth
line leads to the even more bizarre time dependence
$$            \sin(\dm(\tau_1 + \tau_2) -\phi')
                       e^{-(\gb +\alpha-\gamma)(\tau_1+\tau_2)}\ .
\eqn\newtimetwo$$
If we interpret frequencies as energies, both of the forms
 \newtimeone\ and  \newtimetwo \ signal a switch from positive
to negative values of the energy.  This is a rather subtle
 breakdown of energy
conservation, which should be expected in the framework of
density matrix evolution equations according to the analysis of
 ref. \bps.  This subtle effect does not obviously  lead to
macroscopic violations of energy conservation.
However, in the \phif\ experiments, one does not need to wait for the
problems of energy conservation to built up to a macroscopic violation;
one can instead track these violations directly in the frequency
dependence of corrections.
Finally, the second term in the fourth line contains the time-dependences
$$        e^{-\gs(\tau_1 + \tau_2)} \ , \qquad
        e^{-\gl(\tau_1 + \tau_2)} \ .
\eqn\newtimethree$$
Both terms in this line signal a breakdown of the antisymmetry of the
final state wave function, which corresponds to a subtle breakdown of
angular momentum conservation.

  The basic observables computed from $P$ are double differential
decay rates, the probabilities that the kaon with momentum $p$
decays into the final state $f_1$ at proper time $\tau_1$ while the
kaon with momentum $(-p)$ decays to the final state $f_2$ at
proper time $\tau_2$.  We denote this quantity as  $\P(f_1,\tau_1;
f_2,\tau_2)$.  If we denote the expression \Pform \ schematically
as
$$   P = \sum_{i,j}  A_{ij}\, \rho_i \otimes \rho_j \ ,
\eqn\rewriteP$$
where $i,j$ run over $S,L,I, \bar I$, and write the corresponding
eigenvalues as  $\lambda_i$,
then the double decay rate is given by
$$ \P(f_1,\tau_1;f_2,\tau_2) =
    \sum_{i,j}  A_{ij} \, \tr[\rho_i \O_{f_1}] \tr[\rho_j \O_{f_2}]
             e^{-\lambda_i \tau_1 - \lambda_j \tau_2} \ .
\eqn\Peval$$

  Since it is easier to understand a distribution in one variable, much
of the analysis of \phif\ experiments has made use of the integrated
distribution at fixed time interval $\Dt = \tau_1-\tau_2$.  We will
assume, in working with this quantity, that $\Dt > 0$.  Then this
time interval distribution is defined as
$$   \PP(f_1;f_2;\Dt) = \int^\infty_{\Dt} d(\tau_1 + \tau_2) \,
 \P(f_1,\tau_1;f_2,\tau_2)\ .
\eqn\Pdiff$$
This time interval distribution is very useful for obtaining the
standard \cpv\ parameters of the neutral kaon system.  However,
since \Pdiff\ integrates out one of the exponentials in \Peval, this
integral does not possess the strange time dependences which signal
quantum mechanics violation.  A different quantity which shows
these unusual effects more clearly is the double decay rate
interpolated to equal times:
$$   \Q(f_1;f_2;\tau) = \P(f_1,\tau;f_2,\tau) \ .
\eqn\Qcaldef$$
In the case in which quantum mechanics is exact and the density matrix
is given by \phidensitytime, this expression is a linear combination of
decreasing exponentials, with no oscillatory terms.  However, when
we introduce the EHNS parameters, $\Q(\tau)$ can acquire terms
with the oscillatory dependences $\cos(\dm \Dt-\phi)$ and $\cos(2
\dm \Dt - \phi')$.

   The most striking example of this modification of the quantum
mechanical prediction appears in the case of decay to identical
final states $f_1 = f_2 = f$.
  In this case, the quantum mechanical prediction
for the double decay rate is especially simple.\refmark\pecceione
Using
$$  \tr[\rho_S \O_f] \tr[\rho_L \O_f] = \tr[\rho_I\O_f]\tr [ \rib\O_f]
  = \bigl|\VEV{f \bigm| K_S}\bigr|^2 \bigl|\VEV{f \bigm| K_L}\bigr|^2
 \ ,
\eqn\rhotracesf$$
we can rewrite the expression for the double decay rate as
$$  \P^\qmg(f,\tau_1;f,\tau_2) =  C \times \bigl[
      e^{-\gs\tau_1-\gl\tau_2} + e^{-\gl\tau_1 - \gs\tau_2}
  - 2 \cos(\dm (\tau_1-\tau_2)) e^{-\gb(\tau_1+\tau_2)}\bigr]\ .
\eqn\doubleqm$$
 This quantity depends on the two times in a manner completely
fixed by quantum mechanics irrespective of the properties of the
decay amplitudes.  In particular, at $\tau_1 = \tau_2$, the
double distribution {\it vanishes}, as a consequence of the antisymmetry
of the final state wavefunction.  All of these conclusions hold
whether or not \cpt\ symmetry is preserved.

    On the other hand, when quantum mechanics is violated, decays to
identical final states can have a much less constrained structure which
includes all of the time dependences found in the most general case.
We will see examples of this in the next section.

\chapter{Tests of \qm\ at a \phif: identical final states}

   Now that we have clarified the general form of the effects of
quantum mechanics violation in the EHNS model, we will present
expressions for the dependence of particular observables on the
EHNS parameters. In particular, we will  show explicitly how the
terms of the
equal time distributions $\Q(f;f;\tau)$, defined in eq.
 \Qcaldef, can be used to constrain these parameters.
In this section, we will return to the framework of Section 4, in which
\cptv\ in decay vertices and  $\ek'$ are neglected. However, in
contrast to the case of experiments on single kaons, the constraints
on the EHNS parameters from
\phif\ experiments are not  essentially affected by the inclusion of
\cptv\ in decay vertices.  In the next section, we will explain this
point and also discuss some aspects of the measurement of these
decay parameters.

   Consider first the case in which both kaons decay semileptonically.
The cases in which both kaons decay to $\pi^-\ell^+\nu$ or to
$\pi^+\ell^-\bar\nu$ are examples of decays to identical final states
whose special properties were discussed at the end of the previous
section.  It is straightforward to work out the double time
distribution for these cases by using the expression \Pform, the
explicit forms of the density matrix eigenmodes \fourmodestwo, and the
decay operators \Oellplus, \Oellminus.  We find, to first order in
small parameters
$$ \eqalign{
 \P(&\ell^\pm,\tau_1;\ell^\pm,\tau_2) =
    {|a|^4\over 8} \cr
\times\bigg\{ & (1 \pm 4\Re\, \ek_M)\bigl[e^{-\gs \tau_1 - \gl \tau_2} +
e^{-\gl \tau_1 - \gs \tau_2} - 2 \cos(\dm(\tau_1-\tau_2) e^{-(\gb +
\alpha- \gamma)(\tau_1 + \tau_2)}\bigr]\cr
&\pm 4{\beta\over |d|}\sin(\dm \tau_1 - \phi_{SW})e^{-(\gb +
\alpha-\gamma)\tau_1 }e^{-\gs\tau_2} + (1 \leftrightarrow 2)\cr
&\pm 4{\beta\over |d|}\sin(\dm \tau_1 + \phi_{SW})e^{-(\gb +
\alpha-\gamma)\tau_1 }e^{-\gl\tau_2} + (1 \leftrightarrow 2)\cr
& +2 {\alpha\over \dm} \sin\dm(\tau_1+\tau_2) \,  e^{-(\gb + \alpha -
\gamma)(\tau_1 + \tau_2)}
 + 2 {\gamma\over \dg}\bigl[ e^{-\gl(\tau_1 + \tau_2)} -
 e^{-\gs(\tau_1+\tau_2)}\bigr]\bigg\}.\cr}
 \eqn\bigexpll$$
 Notice that the first term in the brackets has a form quite close to
 the canonical form \doubleqm\ predicted by quantum mechanics, while
 the remaining terms give systematic corrections to this result. For
 comparison,
 $$ \eqalign{
 \P(\ell^\pm,\tau_1;\ell^\mp,&\tau_2) =
    {|a|^4\over 8} \cr
\times\bigg\{ & \bigl[e^{-\gs \tau_1 - \gl \tau_2} +
e^{-\gl \tau_1 - \gs \tau_2} + 2 \cos(\dm(\tau_1-\tau_2))
 e^{-\gb(\tau_1 + \tau_2)}\bigr]\cr
 & + \O(\ek,\alpha,\beta,\gamma) \bigg\}.\cr}
 \eqn\littleexpll$$
 The complete expression is given in Appendix B.

  The form of the corrections to quantum mechanics are easiest to see
by interpolating to the line $\tau_1 = \tau_2$. On this line,
\littleexpll\ becomes
$$  \P(\ell^\pm,\tau;\ell^\mp,\tau) =
    {|a|^4\over 8} \cdot 4 e^{-2\gb\tau}\ ,
 \eqn\simplelittlell$$
 and \bigexpll\ has a similar, though less dramatic, simplification.
Then one finds
$$  \eqalign{
\Q(\ell^\pm;\ell^\pm;\tau)&/\Q(\ell^\pm;\ell^\mp;\tau) = \cr
& {1\over2}\bigl[1- e^{-2(\alpha-\gamma)\tau}
\bigl(1-{\alpha\over \dm}\sin
2\dm\tau\bigr)\bigr]\cr
+  & {1\over2}{\gamma\over \dg}\bigl[ e^{+\dg\tau}
- e^{- \dg\tau}\bigr]\cr
\pm  & 2 {\beta\over |d|}\bigl[\sin( \dm\tau-\phi_{SW})e^{-\dg\tau/2} +
\sin( \dm\tau+\phi_{SW})e^{+\dg\tau/2}\bigr]\cr}
\eqn\Qratll$$

The three coefficients $\alpha$, $\beta$, and $\gamma$ are selected by
terms which are monotonic in $\tau$, oscillatory with frequency
$\dm$,and oscillatory with frequency $2\dm$.

It is amusing to note
that the positivity of the expression \Qratll\ under the conditions
\abcrest\ is maintained by a delicate interplay of the correction
terms.  This is most easily seem by examining the limits $\alpha \gg
\gamma > 0$ and $\gamma\gg \alpha > 0$.

Similar information is provided by the decay distribution to
$\pi\pi$ final states on both sides of the $\phi$ decay process.
Since we ignore $\ek'$ effects, the decay distributions to $\pi^+\pi^-$
and $\pi^0\pi^0$ final states are identical up to the overall
normalization.  For the specific case of $\pi^+\pi^-$ decays on both
sides, we find, to second order in small parameters
 $$ \eqalign{
 \P(\pi^+\pi^-,\tau_1;&\pi^+\pi^-,\tau_2) =  {2|A_0|^4} \cr
\times\bigg\{ & R_L
\bigl[e^{-\gs \tau_1 - \gl \tau_2} +
e^{-\gl \tau_1 - \gs \tau_2}\bigr]\cr
 &- 2  |\bar\eta_{+-}|^2\cos(\dm(\tau_1-\tau_2)) e^{-(\gb +
\alpha- \gamma)(\tau_1 + \tau_2)}\cr
&+ 4{\beta\over |d|}|\bar \eta_{+-}|
\sin(\dm \tau_1 +\phi_{+-}- \phi_{SW})
e^{-(\gb + \alpha-\gamma)\tau_1 }e^{-\gs\tau_2}
+ (1 \leftrightarrow 2)\cr
&- 2\biggl[ \ {\gamma\over \dg}+ 2{\beta\over |d|}|\bar\eta_{+-}|
{\sin\phi_{+-}\over
\cos\phi_{SW}}\ \biggr] e^{-\gs(\tau_1+\tau_2)}\bigg\}.\cr}
 \eqn\bigexppp
 $$
where $|\bar\eta_{+-}|e^{i\phi_{+-}} = \ek_L^- $ and $R_L$ is defined
as in \obslongvals. Specializing to the line $\tau_1= \tau_2= \tau$,
we find
 $$ \eqalign{
 \Q(\pi^+\pi^-;&\pi^+\pi^-;\tau) \propto  e^{-2\gb\tau}\cr
  \times\bigg\{&|\bar\eta_{+-}|^2\bigl[ 1 - e^{-2(\alpha -
  \gamma)\tau}\bigr] - 2{\beta\over |d|} |\bar\eta_{+-}|
  {\sin(\phi_{+-}-2\phi_{SW})+e^{-\dg\tau} \sin\phi_{+-}
  \over \cos\phi_{SW} } \cr
  &+ {\gamma\over \dg}\bigl[ 1 - e^{-\dg \tau}\bigr]\cr
 &+ 4{\beta\over |d|}|\bar\eta_{+-}|\sin(\dm\tau + \phi_{+-} - \phi_{SW})
           e^{-\dg\tau/2}\bigg\}+ {\cal
           O}^3(\alpha,\beta,\gamma,\epsilon_M, \Delta). \cr}
 \eqn\Qforpipi$$
 This distribution is less sensitive to $\alpha$; its leading $\alpha$
 effect is of order $\alpha\cdot |\bar\eta_{+-}|^2$.  However, the
 measurement of this distribution does allow one to put independent
 constraints on $\beta$ and $\gamma$.

The corresponding distributions for the three pion decay channels can be
obtained by tracing the density matrix $P$ with the operator $
\O_{3\pi} \otimes  \O_{3\pi} $ as in \Peval. $\O_{3\pi}$ is expressed
in terms of the three pion decay amplitudes as
 $$  \O_{3\pi}   =  |X_{3\pi}|^2
          \pmatrix{ |Y_{3\pi}|^2 &  Y^*_{3\pi} \cr Y_{3\pi} &1 \cr}\ ,
\eqn\Oppthree
 $$
where
$$   X_{3\pi} = \VEV{3\pi \Bigm| K_2 } \ ,\qquad
Y_{3\pi} =   { \VEV{3\pi \bigm| K_1} \over\VEV{3\pi \bigm| K_2}} .
\eqn\Oppdefsthree$$

In this section we ignore \cp\ and \cpt\ violation in the decay
amplitudes, so we set $Y_{3\pi}$ to zero; then
 $$ \eqalign{
 \P(3\pi,\tau_1;&3\pi,\tau_2) =  {|X_{3\pi}|^4 \over 2} \cr
\times\bigg\{ & R_S
\bigl[e^{-\gs \tau_1 - \gl \tau_2} +
e^{-\gl \tau_1 - \gs \tau_2}\bigr]\cr
 &- 2  |\bar\eta_{3\pi}|^2\cos(\dm(\tau_1-\tau_2)) e^{-(\gb +
\alpha- \gamma)(\tau_1 + \tau_2)}\cr
&+ 4{\beta\over |d|}|\bar \eta_{3\pi}|
\sin(\dm \tau_1 -\phi_{3\pi}+ \phi_{SW})
e^{-(\gb + \alpha-\gamma)\tau_1 }e^{-\gl\tau_2}
+ (1 \leftrightarrow 2)\cr
&+ 2 \biggl[{\gamma\over \dg} + 2 {\beta \over |d|}|\bar\eta_{3\pi}|
{\sin\phi_{3\pi} \over \cos\phi_{SW}}\biggr]
e^{-\gl(\tau_1+\tau_2)}\bigg\}+
{\cal O}^3(\alpha,\beta,\gamma,\epsilon_M, \Delta) .\cr}
 \eqn\bigexpthreep$$
where $|\bar\eta_{3\pi}|e^{i\phi_{3\pi}} = \ek_S^+ $ and
$R_S = {-\gamma/\dg}+|\bar\eta_{3\pi}|^2
-4(\beta/ \dg) \Im\bigl[ \bar\eta_{3\pi} d/d^*\bigr]$.

This distribution reduces, on the line $\tau_1= \tau_2= \tau$,
to
$$ \eqalign{
 \Q(3\pi;&3\pi;\tau) \propto  e^{-2\gb\tau}\cr
  \times\bigg\{&|\bar\eta_{3\pi}|^2 \bigl[ 1 - e^{-2(\alpha -
  \gamma)\tau}\bigr] - 2{\beta\over |d|} |\bar\eta_{3\pi}|
  {\sin(2\phi_{SW}-\phi_{3\pi}) - e^{\dg\tau} \sin\phi_{3\pi}
  \over \cos\phi_{SW} } \cr
  - {\gamma\over \dg}&\bigl[ 1 - e^{\dg \tau}\bigr]\cr
   + 4{\beta\over |d|}&|\bar\eta_{3\pi}|\sin(\dm\tau -
  \phi_{3\pi} + \phi_{SW})
           e^{\dg\tau/2}\bigg\}+
  {\cal O}^3(\alpha,\beta,\gamma,\epsilon_M, \Delta)
           . \cr}
 \eqn\Qforpipipi$$

 This distribution has a sensitivity to $\gamma$ which survives at
 large time ($\gs\tau\gg 1$) as
   $$
  \lim_{\gs\tau\gg 1}\,\,\Q(3\pi;3\pi;\tau) \propto {\gamma \over
  \dg} e^{-2\gl\tau}\, .
  \eqn\largetimethreepion
   $$ Its origin can be traced to the $\rl \otimes \rl$ propagating
   mode of the
   density matrix \Pform.
  This behavior contrasts with the behavior of the equal-time
 distribution  $\Q(\pi^+\pi^-;\pi^+\pi^-;\tau)$; the latter
  vanishes, at large time, as $\sim e^{-\gs\tau}$.

\chapter{Tests of \qm\ at a \phif: general final states}

In the previous section, we showed how to constrain the EHNS
parameters independently of one another by considering the most
straightforward experiments on $\phi$ decay, analyzed with the simplest
theory.  In this section, we will generalize our analysis to include
direct \cptv\ and the effects of $\ek'$.
in the $K^0$
decay matrix elements, as we have discussed for single kaon experiments
in Section 5.
We will first re-examine the determination of the EHNS parameters.
We will show that
these complications have virtually no effect on the method, or even the
formulae, given in the previous section for the determination of
$\alpha$, $\beta$, and $\gamma$.  Then we will consider the reciprocal
problem of the effect of quantum mechanics violation on the experimental
determination of the kaon decay matrix elements.  We will
show that the measurements of these decay matrix elements can be
affected if $\beta$ and $\gamma$ are nonzero.  The measurements
most sensitive to the modifications of kaon decay within quantum
mechanics are asymmetries of the integrated time
distributions \Pdiff.\refmark\pecceione  We will present formulae
which show how these asymmetries are shifted by quantum mechanics
violation.

The modification of the formulae for the $\Q(f;f;\tau)$ in the presence
of the effects discussed in Section 5 is  quite minor.  In the
formulae for leptonic double decay distributions, $|a|^2$ is changed to
$|a+b|^2$ for decays to $\ell^+$ and to $|a-b|^2$ for decays to
$\ell^-$.  This has no effect on the functional form of the
double time distribution except for the simple replacement
$$\ek_M \rightarrow \ek_M + {b \over a}\, .
\eqn\changeepsm
$$
In particular, the formula \Qratll\ is still
valid in this more general context.  Similarly, the inclusion of more
general effects in the kaon decay vertices changes the relative
normalization of $\pi^+\pi^-$ and $\pi^0\pi^0$ decay rates.  However,
the formula \Qforpipi\ remains valid with the replacement
$$ |\bar\eta_{+-}| e^{i\phi_{+-}} =    \ek_L^-  + Y_{+-}\ ,
\eqn\changetwo$$
as in eq. \changethetimedefs. This is equally true for the case of
the three pion decay, for which  formula \Qforpipipi\ remains valid
with the substitution\foot{We also assume here that we can neglect
a term proportional to $\Im Y_{3\pi}$; we expect this term
to be no bigger
than $\Im Y_{+-} \sim \Im \epsilon'$.}
$$ |\bar\eta_{3\pi}| e^{i\phi_{3\pi}} =    \ek_S^+  + Y_{3\pi}\ .
\eqn\changethree$$
  Thus, there is no difficulty in
constraining \cptv\ from outside quantum mechanics in \phif\
experiments even in this more general context.  For reference, the
complete expressions for the double time distributions to pion or
leptonic final states are given in Appendix B.

It is also interesting to ask whether the converse of this statement is
true, whether quantum mechanics violation can interfere with the
measurement of $\ek'$ and other more standard asymmetries which should
appear at a \phif. In the following, we briefly discuss how the
determination of certain of these quantities might be
affected.

One would not immediately expect the
 measured value of $\ek'/\ek$ to be significantly
affected by quantum mechanics violation,  since $\ek'$ is a
property of the difference between the decay rates into
 $\pi^+\pi^-$ and
$\pi^0\pi^0$ while the parameters $\beta$ and $\gamma$
affect the time evolution of the kaon
system prior its decay. From this argument,
 one would expect $\ek'/\ek$ to be
at most corrected by a factor  $ 1+ {\cal
O}(\,\beta/d\ek\,,\,\gamma/|d||\ek|^2)\, $. Using the bounds
on $\beta$ and $\gamma$  derived in Section 4, we
estimate the terms in parenthesis to be at most corrections
of the order of
$5\%$ and $25\%$ respectively. These corrections are mild and can be
reduced further if no evidence for nonzero $\beta$ and $\gamma$
is found at a \phif.

However, if $\beta$ and $\gamma$ prove to be nonzero, they will have
two important effects in the measurement of $\ek'/\ek$.  First,
any time-dependent method of determining
$\ek'/\ek$ would be complicated by the new time dependences introduced
by quantum-mechanics violating terms in the density matrix evolution.
As we have already pointed out, this effect can
be minimized by considering the integrated distributions
\Pdiff\ at fixed time
interval $\Dt=\tau_1-\tau_2$.
Second, because $\beta/d$ and $\ek$ are close to being
orthogonal in the complex
plane, effects of nonzero $\beta$ can mix
the real and imaginary parts of $\ek'/\ek$.

 To demonstrate these points quantitatively, we
consider the determination\refmark\dunietz of $\ek'/\ek$
from the measurement of the quantity
$$
{\cal A}( \pi^+\pi^-;\pi^0\pi^0;\Dt)=
{\PP(\pi^+\pi^-;\pi^0\pi^0;\Dt)- \PP(\pi^0\pi^0;\pi^+\pi^-;\Dt) \over
\PP(\pi^+\pi^-;\pi^0\pi^0;\Dt)+ \PP(\pi^0\pi^0;\pi^+\pi^-;\Dt) }\, .
\eqn\ppepsprime
$$
This asymmetry can be
 straightforwardly computed using the formulae of Appendix B.
We find
$$
{\cal A}( \pi^+\pi^-;\pi^0\pi^0;\Dt)= 3 \Re \, {\ek'\over\ek}
\times{ {\cal
N}_R \over {\cal D}}\, - 3\, \Im\,{\ek'\over\ek} \times { {\cal
N}_I \over {\cal D}}
 \, ,
\eqn\ppeps
$$
The coefficients ${\cal N}_{R,I}$ and ${\cal D}$ are  functions
of $\Dt$, $\beta/d$, $\gamma$, and $|\etapp|$ which are given
in Appendix C.

In the context of pure \qm, the quantities ${\cal N}_{R,I}$ and
${\cal D}$ have a simple functional form, and the
quantities $\Re\,\ek'/\ek $, $\Im\,\ek'/\ek $ can be
extracted from ${\cal A}^\qmg( \pi^+\pi^-;\pi^0\pi^0;\Dt)$ by a
 two-parameter fit. In presence of quantum mechanics violation,
this is no longer true. For example, in the limit $\gs\Dt \gg 1$,
$${\cal A}^\qmg( \pi^+\pi^-;\pi^0\pi^0;\Delta \tau) \rightarrow
3\Re \, \ek'/\ek \ .
\eqn\qmgAform$$
In the same limit, $\ppeps$ becomes
$$ \eqalign{
{\cal A}( \pi^+\pi^-&;\pi^0\pi^0; \Dt)  \rightarrow   \cr
&3 \Re \, \ek'/\ek\,\,\biggl[{
1+ (2\beta/|d||\etapp|) \sin (  \phi_{SW}-\phipp )
\over 1+ (\gamma/ \dg |\etapp|^2) + (2\beta/
|d||\etapp|)
 (\sin ( 2 \phi_{SW}-\phipp )/\cos \phi_{SW})  }   \biggr]
\cr
-&3\Im \, \ek'/\ek\,\,\biggl[ (2\beta/|d||\etapp|)\, \cos(\phi_{SW}
-\phipp)\biggr] \cr
} \, .
\eqn\eprimexact$$
To understand the role of the
near-orthogonality of $\beta/d$ and $\etapp$,
we may approximate $\phipp \approx
\phi_{SW} \approx 45^\circ$; then \eprimexact\ simplifies to
$$\eqalign{
{\cal A}( \pi^+\pi^-;&\pi^0\pi^0; \Dt)  \rightarrow  \cr
 &
3\Re \, \ek'/\ek\,\,\biggl[
1- {\gamma \over{\sqrt 2} |d| |\etapp|^2} - 2{\beta\over
|d||\etapp|}     \biggr]
-3\,\Im \, \ek'/\ek\,\,\biggl[ 2{\beta\over |d||\etapp|}\biggr] \cr }\,
{}.
\eqn\eprimeapprox$$
This makes clear that, if one opens the possibility of quantum mechanics
violation in the neutral kaon system at the level
$\beta \sim m_K^2/m_{\rm Pl}$,
one must constrain or
determine $\beta$ in order to measure $\Re\, \ek'$.

 Similar effects are present in asymmetries involving the
semileptonic decay distributions
 $\PP(\ell^\pm;\ell^\pm;\Dt)$.  For simplicity, we again consider
only the large-time limit $\gs\Dt \gg 1$.  Three particularly
informative asymmetries are
$$\eqalign{
{\PP(\ell^+;\ell^+;\Dt)-\PP(\ell^-;\ell^+;\Dt)
\over
\PP(\ell^+;\ell^+;\Dt)+\PP(\ell^-;\ell^-;\Dt) } &\ra
\delta_L
\cr
{\PP(\ell^+;\ell^+;\Dt)-\PP(\ell^-;\ell^-;\Dt)
\over
\PP(\ell^+;\ell^+;\Dt)+\PP(\ell^-;\ell^-;\Dt) } &\ra
4\Re\,(\ek_M + b/a)\, + \,8 {\beta \over |d|}\sin 2\phi_{SW}\,
\cos\phi_{SW}
\cr
{\PP(\ell^+;\ell^-;\Dt)-\PP(\ell^-;\ell^+;\Dt)
\over
\PP(\ell^+;\ell^+;\Dt)\,+\,\PP(\ell^-;\ell^-;\Dt) } &\ra
-4\Re\,\Delta + 4 {\beta \over |d|}\sin \phi_{SW}\bigl[ 1 - 4 \cos^2
\phi_{SW} \bigr] \ .
\cr}
\eqn\integratedell$$
The first of the above formulae yields a direct determination of
$\delta_L$, even in the presence of quantum mechanics violation.
However, the other two formulae are more complicated.   If quantum
mechanics is assumed to be
valid, these two limits give simple constraints on the
\cpt\ violating parameters $\Delta$ and $\Re(b/a)$.  However, in
the more general context of quantum mechanics violation, these
parameters are constrained only to the extent that $\beta$ is known
from the experiments described in Section 7.

   It is remarkable how sensitively the parameters of quantum
mechanics violation affect the various observable quantities of the
$K^0$--$\bar K^0$ system as observed at the $\phi$.
Using the strategies we have discussed, it is
 likely that
all three of the EHNS parameters of quantum mechanics violation
can be bounded, or measured, at a
level well  below
$(1/m_{Pl})$.  Perhaps there are still more surprises waiting for us in
neutral kaon physics.

\APPENDIX{A}{A: Eigenmodes of density matrix evolution}

  In this appendix, we give more exact formulae for the eigenmodes of
the density matrix evolution and for the components of the density
matrix for neutral kaon pairs resulting from $\phi$ decay.  The
expressions below are complete through second order in \cp-violating
parameters.

   First, we present the density matrix eigenmodes.  These are given
as column vectors
$$
\pmatrix{\rho_1\cr \rho_2\cr
               \ri\cr \rib\cr} \ ,
\eqn\beginrho$$
where these elements are defined by eq. \densitymat.  The expressions
\fourmodestwo\ should be replaced by:
\def\aaa{-\gamma/\dg + 2\Im[\ek^+_S\ek^{-*}_Sd]/\dg}
\def\bbb{\ek^{-*}_S - (d\ek^+_L\gamma /d^*\dg)
    +i(\ek^{-*}_S\gamma/d^*) -  (2\alpha \Im\, \ek_S^-]/d^*) }
\def\ccc{\ek^{-}_S - (d^*\ek^{+*}_L\gamma /d\dg)
    -i(\ek^{-}_S\gamma/d) -  (2\alpha \Im\, \ek_S^-]/d) }
$$
\rho_S = \pmatrix{
 1 \crr  \aaa\crr\bbb\crr \ccc\cr}  \ ,
\eqn\SSSS$$
\def\aab{\gamma/\dg - 2\Im[\ek^+_L\ek^{-*}_Ld^*]/\dg}
\def\aac{\ek^{+}_L + (d^*\ek^{-*}_S\gamma /d\dg)
    +i(\ek^{+}_L\gamma/d) +  (2\alpha \Im\, \ek_L^+]/d) }
\def\aad{\ek^{+*}_L + (d\ek^{-}_S\gamma /d^*\dg)
    -i(\ek^{+*}_L\gamma/d^*) +  (2\alpha \Im\, \ek_L^+]/d^*) }
$$
\rho_L =
\pmatrix{\aab \crr 1 \crr
               \aac \crr \aad \cr} \ ,
\eqn\LLLL$$
\def\aae{\ek^{-*}_L(1+i\gamma/d^*-i \alpha/d^*) -i\ek_S^+ \gamma/d^* + i
      (d\ek^-_L\alpha/2d^* \dm )  }
\def\aaf{\ek^{+}_S(1+i\gamma/d-i\alpha/d) -i\ek_L^{-*} \gamma/d + i
      (d^*\ek^{+*}_S\alpha/2d \dm )  }
\def\aag{-i (\alpha/2\dm) +
      (\ek^-_S \ek^{-*}_L d + \ek^{+*}_L\ek^{+*}_S d^*)/2\dm}
$$
\rho_I =  \pmatrix{\aae \crr  \aaf \crr
          1 \crr  \aag \cr} \ ,
\eqn\IIII$$
\def\aah{\ek^{-}_L(1-i\gamma/d+i \alpha/d) +i\ek_S^{+*} \gamma/d^* - i
      (d^*\ek^{-*}_L\alpha/2d \dm )  }
\def\aai{\ek^{+*}_S(1-i\gamma/d^*+i\alpha/d^*) +i\ek_L^{-} \gamma/d^* - i
      (d\ek^{+}_S\alpha/2d^* \dm )  }
\def\aaj{i (\alpha/2\dm) +
       (\ek^{-*}_S\ek^-_Ld^*+\ek^+_L\ek^{+*}_Sd)/2\dm}
$$
\rib = \pmatrix{\aah \crr \aai \crr \aaj \crr  1\cr} \ .
\eqn\IBBB$$
Note that the expressions for $\rho_{L1}$ and $\rho_{S2}$ given in
\fourmodestwo\ agree with the expressions above.

Next, we present a more precise form for the density matrix $P$ by
quoting the matrix elements of
$A_{ij}$, defined by eq. \rewriteP, to second order in small parameters.
We find:
$$ \eqalign{
A_{SL} &= A_{LS} = {1\over2}(1+2\Re(\ek_S\ek_L))+\beta\Re
{\ek_S-\ek_L\over d} + {3\over2}\, \beta^2 \
{d^2+d^{*2}\over|d|^{4}}-{3\over2}\ {\gamma^2\over(\dg)^2}  \crr
A_{SS} &= - {\gamma\over \dg}-2\ {\beta^2\over|d|^2}- 4\,
{\beta\over \dg}\, \Im \ek_L \crr
A_{LL} &= {\gamma\over\dg}-2\, {\beta^2\over|d|^2}+ 4\, {\beta\over\dg}\,
\Im \ek_S \cr
A_{SI} &= A_{IS} = (A_{S\overline I})^* = (A_{\overline IS})^*
\crr
&= -{\beta\over d}-2\, {\alpha\over d}\ \Im\ek_L-i\, {\gamma\over d}
\, (\ek_L-\ek^*_S) + i\, {\alpha\beta\over 2\dm\, d^*}-
{\gamma\beta\over \dg\, d^*}\crr
A_{LI} &= A_{IL} = (A_{L\overline I})^*=(A_{\overline IL})^*
\crr &=  {\beta\over d^*}+ 2\, {\alpha\over d^*}\, \Im\ek_S+i\,
{\gamma\over d^*}\, (\ek_L-\ek^*_S)-i\, {\alpha\beta\over 2\dm d}
- {\beta\gamma\over \dg d} \crr
A_{I\bar I} &= A_{\bar I I} = -{1\over2}\, (1+2\Re
\ek_L\ek_S)+\beta\, \Re \left(\ek_S-\ek_L\over d\right) \crr
 &\hskip 0.4in  + {3\over 2}\, \beta^2\, {d^2+(d^*)^2\over |d|^{4}}
- {3\over8}\ {\alpha^2\over (\dm)^2}\crr
A_{II} &= (A_{\overline I\overline I})^* = {i\alpha\over 2\dm} - 2\,
{\beta^2\over |d|^2}-{\beta\over\dm}\, (\ek_L-\ek^*_S)\cr}
\eqn\AAAAA$$

\APPENDIX{B}{B: Double time distributions}

  In this appendix, we give the complete formulae for the double time
  distributions $\P(f_1,\tau_1;f_2,\tau_2)$ for $\phi$ decay to the
  various final states discussed in the text.  The
expressions below are complete through first order in \cp-violating
parameters unless it is specified otherwise.

$$ \eqalign{
 \P(\ell^\pm,&\tau_1;\ell^\pm,\tau_2) =
    {|a|^4\over 8} \cr
\times\bigg\{ \bigl[ (1 \pm 4\Re\,& (\ek_M + {b\over a} )\bigr]
\bigl[e^{-\gs \tau_1 - \gl \tau_2} +
e^{-\gl \tau_1 - \gs \tau_2} - 2 \cos(\dm(\tau_1-\tau_2)) e^{-(\gb +
\alpha- \gamma)(\tau_1 + \tau_2)}\bigr]\cr
&\pm 4{\beta\over |d|}\sin(\dm \tau_1 - \phi_{SW})e^{-(\gb +
\alpha-\gamma)\tau_1 }e^{-\gs\tau_2} + (1 \leftrightarrow 2)\cr
&\pm 4{\beta\over |d|}\sin(\dm \tau_1 + \phi_{SW})e^{-(\gb +
\alpha-\gamma)\tau_1 }e^{-\gl\tau_2} + (1 \leftrightarrow 2)\cr
& +2 {\alpha\over \dm} \sin\dm(\tau_1+\tau_2) \,  e^{-(\gb + \alpha -
\gamma)(\tau_1 + \tau_2)}
 + 2 {\gamma\over \dg}\bigl[ e^{-\gl(\tau_1 + \tau_2)} -
 e^{-\gs(\tau_1+\tau_2)}\bigr]\bigg\}.\cr}
 \eqn\bigexpll$$

$$ \eqalign{
 \P(\ell^+,\tau_1;&\ell^-,\tau_2) =
    {|a|^4\over 8} \cr
\times\bigg\{ & (1 +4\Re\,(\Delta - \beta/d))
e^{-\gs \tau_1 - \gl \tau_2} +
(1 -4\Re\,(\Delta - \beta/d))e^{-\gl \tau_1 - \gs \tau_2} \cr
 &+ 2 \cos\bigr(\dm(\tau_1-\tau_2)-
 4\Im(\Delta+\beta/d)\bigl)\, e^{-(\gb +
\alpha- \gamma)(\tau_1 + \tau_2)}\cr
&+ 4{\beta\over |d|}\sin(\dm \tau_1 - \phi_{SW})e^{-(\gb +
\alpha-\gamma)\tau_1 }e^{-\gs\tau_2} - (1 \leftrightarrow 2)\cr
&+ 4{\beta\over |d|}\sin(\dm \tau_1 + \phi_{SW})e^{-(\gb +
\alpha-\gamma)\tau_1 }e^{-\gl\tau_2} - (1 \leftrightarrow 2)\cr
& -2 {\alpha\over \dm} \sin\dm(\tau_1+\tau_2) \,  e^{-(\gb + \alpha -
\gamma)(\tau_1 + \tau_2)}
 + 2 {\gamma\over \dg}\bigl[ e^{-\gl(\tau_1 + \tau_2)} -
 e^{-\gs(\tau_1+\tau_2)}\bigr]\bigg\}.\cr}
 \eqn\bigexpllm$$

$$ \eqalign{
 \P(\pi^+\pi^-,&\tau_1;\pi^+\pi^-,\tau_2) =
    {|X_{+-}|^4\over 2} \cr
\times\bigg\{  R_L\bigl[&e^{-\gs \tau_1 - \gl \tau_2}
 + e^{-\gl \tau_1 - \gs \tau_2}\bigr]
 - 2|\bar\eta_{+-}|^2 \cos(\dm(\tau_1-\tau_2))\, e^{-(\gb +
\alpha- \gamma)(\tau_1 + \tau_2)}\cr
&+ 4{\beta\over |d|}|\bar\eta_{+-}|
\sin(\dm \tau_1+ \phi_{+-} - \phi_{SW})e^{-(\gb +
\alpha-\gamma)\tau_1 }e^{-\gs\tau_2} + (1 \leftrightarrow 2)\cr
& - 2 \biggl({\gamma\over
\dg}+4{\beta\over \dg}\Im \bigr[\bar\eta_{+-}-Y_{+-} \bigr] \biggr)
e^{-\gs(\tau_1+\tau_2)}\bigg\} \ + \  {\cal O}^3(\alpha,\beta,\gamma,
\epsilon_{S,L},Y_{+-}) \ ,\cr}
 \eqn\bigexpllma$$
using $\bar\eta_{+-}e^{i\phi_{+-}} = \ek_L^- + Y^{+-}$
and
$$R_L = {\gamma/\dg}+|\bar\eta_{+-}|^2
+4(\beta/ \dg) \Im\bigl[ \bar\eta_{+-} d/d^*- Y_{+-}   \bigr]\ .
\eqn\RLagain$$

$$ \eqalign{
 \P(\pi^+&\pi^-,\tau_1;\pi^0\pi^0,\tau_2) =
    {|X_{+-}|^2|X_{00}|^2 \over 2} \cr
\times\bigg\{ & R_L^{00} e^{-\gs \tau_1 - \gl \tau_2} +
R_L^{+-} e^{-\gl \tau_1 - \gs \tau_2} \cr
 &- 2 |\bar\eta_{+-}||\bar\eta_{00}| \cos(\dm(\tau_1-\tau_2)
 + \phi_{+-} - \phi_{00})\,
  e^{-(\gb + \alpha- \gamma)(\tau_1 + \tau_2)}\cr
&+ 4{\beta\over |d|}|\bar\eta_{+-}|\sin(\dm \tau_1
- \phi_{SW} +\phi_{+-})e^{-(\gb +
\alpha-\gamma)\tau_1 }e^{-\gs\tau_2} \cr
&+ 4{\beta\over |d|}|\bar\eta_{00}|\sin(\dm \tau_2
- \phi_{SW} +\phi_{00})e^{-(\gb +
\alpha-\gamma)\tau_2 }e^{-\gs\tau_1} \cr
& - 2 \biggl({\gamma\over \dg} + 2{\beta\over\dg}
\Im\bigl[\bar\eta_{+-}-Y_{+-}+\bar\eta_{00}-Y_{00}\bigr] \biggr)
e^{-\gs(\tau_1+\tau_2)}\bigg\} \ + \ {\cal O}^3(\alpha,\beta,\gamma,
\epsilon_{S, L},Y_{\{{+- \atop 00}})\ .\cr}
 \eqn\bigexpllm$$

 $$ \eqalign{
 \P(3\pi,\tau_1;&3\pi,\tau_2) =  {|X_{3\pi}|^4 \over 2} \cr
\times\bigg\{ & R_S
\bigl[e^{-\gs \tau_1 - \gl \tau_2} +
e^{-\gl \tau_1 - \gs \tau_2}\bigr]\cr
 &- 2  |\bar\eta_{3\pi}|^2\cos(\dm(\tau_1-\tau_2)) e^{-(\gb +
\alpha- \gamma)(\tau_1 + \tau_2)}\cr
&+ 4{\beta\over |d|}|\bar \eta_{3\pi}|
\sin(\dm \tau_1 -\phi_{3\pi}+ \phi_{SW})
e^{-(\gb + \alpha-\gamma)\tau_1 }e^{-\gl\tau_2}
+ (1 \leftrightarrow 2)\cr
&+ 2 \biggl( {\gamma\over \dg}+ 4{\beta\over\dg}
\Im\bigl[\bar\eta_{3\pi} - Y_{3\pi}\bigr] \biggr)
e^{-\gl(\tau_1+\tau_2)}\bigg\} \  +  \ {\cal O}^3(\alpha,\beta,\gamma,
\epsilon_{S,L},Y_{3\pi}) .\cr}
 \eqn\bigexpthreep$$
where $|\bar\eta_{3\pi}|e^{i\phi_{3\pi}} = \ek_S^+ + Y_{3\pi}$ and
$$R_S = {-\gamma/\dg}+|\bar\eta_{3\pi}|^2
-4(\beta/ \dg) \Im\bigl[ \bar\eta_{3\pi} d/d^*- Y_{3\pi}\bigr] \, .
\eqn\RSvalue$$

$$ \eqalign{
\P(\pi^+\pi^-,\tau_1;&\ell^{\pm},\tau_2) =
{|X_{+-}|^2\over 2}{|a|^2 \over 2} \cr  \times \bigg\{
&R_Le^{-\gl \tau_1 - \gs \tau_2} +
\bigl[ 1 \pm \delta_L + {\gamma\over \dg} \bigr]e^{-\gs \tau_1 -
\gl \tau_2} \cr
&\mp 2 |\bar\eta_{+-}| \cos(\dm(\tau_1-\tau_2) + \phi_{+-})
e^{-(\gb +\alpha- \gamma)(\tau_1 + \tau_2)}  \cr
&\pm 4{\beta\over |d|}
\sin(\dm \tau_2 - \phi_{SW})e^{-(\gb +
\alpha-\gamma)\tau_2 }e^{-\gs\tau_1} \cr
&- 2 \biggl({\gamma\over \dg} +4{\beta\over\dg}  \Im \,
\bar\eta_{+-} \biggr)
e^{-\gs(\tau_1+\tau_2)}\bigg\}. \cr }
\eqn\asympions
$$

\APPENDIX{C}{C: Formulae for the measurement of $\ek'/\ek$ }
In this Appendix, we present formulae for extracting $\ek'/\ek$
from the integrated distributions at fixed time interval
$\Dt=\tau_1-\tau_2$ of the asymmetric decay into charged and neutral
pion final states.
$$
{\cal A}( \pi^+\pi^-;\pi^0\pi^0;\Dt)=
{\PP(\pi^+\pi^-;\pi^0\pi^0;\Dt)- \PP(\pi^0\pi^0;\pi^+\pi^-;\Dt) \over
\PP(\pi^+\pi^-;\pi^0\pi^0;\Dt)+ \PP(\pi^0\pi^0;\pi^+\pi^-;\Dt) }\, .
\eqn\ppepsprime
$$

$$
{\cal A}( \pi^+\pi^-;\pi^0\pi^0;\Dt)= 3 \Re \, \ek'/\ek \times{ {\cal
N}_R \over {\cal D}}                - 3 \Im \, \ek'/\ek \times{ {\cal
N}_I \over {\cal D}} \, ,
\eqn\ppeps
$$with

$$\eqalign{ {\cal N}_R = e^{-\gl\Dt}& \,\biggl[ 1+
2{\beta\over|d||\etapp|}\sin(\phi_{SW}-\phipp) \biggr]   \cr
                        - e^{-\gs\Dt}& \,\biggl[ 1+
2{\beta\over|d||\etapp|}\Bigl( \sin(\phi_{SW}-\phipp)
-  |z| \sin(\phi_{SW}+\phi_z-\phipp) \Bigr)  \biggr] \cr
                        + e^{-\gb\Dt}& \,\biggl[
2{\beta\over|d||\etapp|}|z|\sin(\dm\Dt+\phipp-\phi_{SW}-\phi_z)
\biggr] \cr }\, ,
\eqn\coeffr
$$

$$\eqalign{ {\cal N}_I = e^{-\gl\Dt}& \,\biggl[
2{\beta\over|d||\etapp|}\cos(\phi_{SW}-\phipp) \biggr] \cr
                        - e^{-\gs\Dt}& \,\biggl[
2{\beta\over|d||\etapp|}\Bigl( \cos(\phi_{SW}-\phipp)
-  |z| \cos(\phi_{SW}+\phi_z-\phipp) \Bigr)  \biggr] \cr
                        + e^{-\gb\Dt}& \,\biggl[2\sin\dm\Dt -
2{\beta\over|d||\etapp|}|z|\cos(\dm\Dt+\phipp-\phi_{SW}-\phi_z)
\biggr] \cr }\, ,
\eqn\coeffi
$$

$$\eqalign{ {\cal D} = e^{-\gl\Dt}& \,\biggl[
1+{\gamma\over\dg|\etapp|^2}+
2{\beta\over|d||\etapp|}{\sin(2\phi_{SW}-\phipp)
\over \cos\phi_{SW}} \biggr]   \cr
                     + e^{-\gs\Dt}& \biggl[
1-{\gamma\over\dg|\etapp|^2}{\gl\over\gs}+
2{\beta\over|d||\etapp|}\Bigl( {\sin(2\phi_{SW}-\phipp)
\over \cos\phi_{SW}}
-2 |z| \sin(\phi_{SW}+\phi_z-\phipp)  \Bigr) \biggr]   \cr
                     - e^{-\gb\Dt}& \,\biggl[ 2 \cos\dm \Dt-
4{\beta\over|d||\etapp|} |z| \sin(\dm\Dt +\phipp-\phi_{SW}-\phi_z)
\biggr]
\cr}
\eqn\coeffd$$

We have defined
$$ |z|e^{i\phi_z} = { 2\gb \over \gs + \gb + i \dm } \, .
\eqn\zzzzzz$$

\ack

We are grateful to Vera L\"uth  and to Paolo and Juliet Franzini
for correspondence
on experimental issues in the neutral kaon system, and to
Philippe Eberhard for valuable  correspondence and conversations.
  We also thank
 our colleagues in the SLAC theory group for many helpful
discussions.

\refout
\endpage
\figout
\bye